\documentclass[11pt]{article}

\usepackage[english]{babel}
\usepackage{amsmath, amssymb, amsfonts}
\usepackage{array}
\usepackage{graphicx}
\usepackage{float}
\usepackage{theorem}
\usepackage[textheight=0.8\paperheight]{geometry}
\usepackage{blindtext}
\usepackage{url}
\usepackage{lscape}
\usepackage{booktabs}
\usepackage{geometry}
\usepackage{times}
\usepackage{bbm}

\renewcommand{\baselinestretch}{1.3}

\newcommand{\Var}{\text{Var}}

\newcommand{\eg}{\textit{e.g.~}}
\newcommand{\etal}{\textit{et~al}.}

\newcommand{\vb}{\mathbf{b}}

\newcommand{\vg}{\mathbf{g}}

\newcommand{\vx}{\mathbf{x}}

\newcommand{\vz}{\mathbf{z}}

\newcommand{\zero}{\mathbf{0}}

\newcommand{\mB}{\mathbf{B}}

\newcommand{\mG}{\mathbf{G}}

\newcommand{\mI}{\mathbf{I}}

\newcommand{\eps}{\epsilon}

\newcommand{\vbeta}{\text{\boldmath{$\beta$}}}

\newcommand{\vgamma}{\text{\boldmath{$\gamma$}}}

\newcommand{\vkappa}{\text{\boldmath{$\kappa$}}}

\newcommand{\mSigma}{\mathbf{\Sigma}}

\newcommand{\AL}{\mathcal{AL}}
\newcommand{\AEP}{\mathcal{AEP}}
\newcommand{\B}{\mathcal{B}}

\newcommand{\DP}{\mathcal{DP}}
\newcommand{\E}{\mathcal{E}}
\newcommand{\G}{\mathcal{G}}
\newcommand{\IG}{\mathcal{IG}}

\newcommand{\N}{\mathcal{N}}
\newcommand{\TN}{\mathcal{TN}}

\newcommand{\SN}{\mathcal{SN}}
\newcommand{\ST}{\mathcal{ST}}

\newcommand{\U}{\mathcal{U}}
\newcommand{\GIG}{\mathcal{GIG}}

\newcommand{\jasa}{\textit{Journal of the American Statistical Association}}
\newcommand{\csda}{\textit{Computational Statistics \& Data Analysis}}
\newcommand{\joe}{\textit{Journal of Econometrics}}
\newcommand{\biometrika}{\textit{Biometrika}}
\newcommand{\econometrica}{\textit{Econometrica}}
\newcommand{\sac}{\textit{Statistics and Computing}}

\newcommand{\jae}{\textit{Journal of Applied Econometrics}}
\newcommand{\jrssb}{\textit{Journal of Royal Statistical Society Series B}}

\begin{document}

\title{{Bayesian Endogenous Tobit Quantile Regression}}

\author{Genya Kobayashi
\thanks{
Chiba University, 
Faculty of Law, Politics, \& Economics. 1-33, Yayoi-cho, Inage-ku, Chiba, 263-8522, Japan. 
Email: \small{gkobayashi@chiba-u.jp}
}\\
}
\date{\today}
\maketitle
\vspace*{0.5em}

\begin{abstract}
This study proposes $p$-th Tobit quantile regression models with endogenous variables. 
In the first stage regression of the endogenous variable on the exogenous variables, the assumption that the $\alpha$-th quantile of the error term is zero is introduced. 
Then, the residual of this regression model is included in the $p$-th quantile regression model in such a way that the $p$-th conditional quantile of the new error term is zero. 
The error distribution of the first stage regression is modelled around the zero $\alpha$-th quantile assumption by using parametric and semiparametric approaches.  
Since the value of $\alpha$ is a priori unknown, it is treated as an additional parameter and is estimated from the data. 
The proposed models are then demonstrated by using simulated data and real data on the labour supply of married women. 

\end{abstract}

\section{Introduction}
\label{Sect:intro}
Since the seminal work of Koenker and Bassett~(1978), quantile regression has received substantial scholarly attention as an important alternative to conventional mean regression. 
Indeed, there now exists a large literature on the theory of quantile regression (see, for example, Koenker~(2005), Yu~\etal~(2003), and Buchinsky~(1998) for an overview).  
Notably, quantile regression can be used to analyse the relationship between the conditional quantiles of the response distribution and a set of regressors, while conventional mean regression only examines the relationship between the conditional mean of the response distribution and the regressors. 

Quantile regression can thus be used to analyse data that include censored responses. 
Powell~(1984; 1986) proposed a Tobit quantile regression (TQR) model utilising the equivariance of quantiles under monotone transformations. 
Hahn~(1995), Buchinsky and Hahn~(1998), Bilias~\etal~(2000), Chernozhukov and Hong~(2002), and  Tang~\etal~(2012)  considered alternative approaches to estimate TQR. 
More recent works in the area of censored quantile regression include Wang and Wang~(2009) for random censoring using locally weighted censored quantile regression, Wang and Fygenson~(2009) for longitudinal data,  Chen~(2010) and Lin~\etal~(2012) for doubly censored data using the maximum score estimator and weighted quantile regression, respectively, and Xie~\etal~(2015) for varying coefficient models.

In the Bayesian framework, Yu and Stander~(2007) considered TQR by extending the Bayesian quantile regression model of Yu and Moyeed~(2001) and proposed an estimation method based on Markov chain Monte Carlo (MCMC). 
A more efficient Gibbs sampler for the TQR model was then proposed by Kozumi and Kobayashi~(2011). 
Further extensions of Bayesian TQR have also been considered.  
Kottas and Krnjaji\'c~(2009)  and Taddy and Kottas~(2012) examined semiparametric and nonparametric models using Dirichlet process mixture models. 
Reich and Smith~(2013) considered a semiparametric censored quantile regression model where the quantile process is represented by a linear combination of basis functions. 
To accommodate nonlinearity in data, Zhao and Lian~(2015) proposed a single-index model for Bayesian TQR. 
Furthermore, Kobayashi and Kozumi~(2012) proposed a model for censored dynamic panel data. 
For variable selection in Bayesian TQR, Ji~\etal~(2012) applied the stochastic search, Alhamzawi and Yu~(2014) considered a $g$-prior distribution with a ridge parameter that depends on the quantile level, and Alhamzawi~(2014) employed the elastic net.

As in the case of ordinary least squares, standard quantile regression estimators are biased when one or more regressors are correlated with the error term. 
Many authors have analysed quantile regression for uncensored response variables with endogenous regressors, such as Amemiya~(1982), Powell~(1983), Abadie~\etal~(2002), Kim and Muller~(2004), Ma and Koenker~(2006), Chernozhukov and Hansen~(2005; 2006; 2008), and Lee~(2007). 

Extending the quantile regression model to simultaneously account for censored response variables and endogenous variables is a challenging issue. 
In the case of the conventional Tobit model with endogenous regressors, a number of studies were published in the 1970s and 1980s, such as Nelson and Olsen~(1978), Amemiya~(1979), Heckman~(1978), and Smith and Blundell~(1986), with more efficient estimators proposed by Newey~(1987) and Blundell and Smith~(1989). 
On the contrary, few studies have estimated censored quantile regression with endogenous regressors. 
While Blundell and Powell~(2007) introduced control variables as in Lee~(2007) to deal with the endogeneity in censored quantile regression, their estimation method involved a high dimensional nonparametric estimation and can be computationally cumbersome. 
Chernozhukov~\etal~(2014) also introduced  control variables to account for endogeneity. 
They proposed using quantile regression and distribution regression (Chernozhukov~\etal,~2013) to construct the control variables and extended the estimation method of Chernozhukov and Hong~(2002).

In the Bayesian framework, mean regression models with endogenous variables have garnered a great deal of research attention from both the theoretical and the computational points of view (\eg Rossi~\etal,~2005; Hoogerheide~\etal,~2007a,~2007b; Conely~\etal,~2008; Lopes and Polson,~2014). 
However, despite the growing interest in and demand for Bayesian quantile regression, the literature on Bayesian quantile regression with endogenous variables remains sparse. 
Lancaster and Jun~(2010) utilised the exponentially tilted empirical likelihood and employed the moment conditions used in Chernozhukov and Hansen~(2006). 
In the spirit of Lee~(2007), Ogasawara and Kobayashi~(2015) employed a simple parametric model using two asymmetric Laplace distributions for panel quantile regression.  
However, these methods are only applicable to uncensored data.  
Furthermore, the model of Ogasawara and Kobayashi~(2015) can be restrictive because of the shape limitation of the asymmetric Laplace distribution, which can affect the estimates. 
Indeed, the modelling of the first stage error in this approach remains to be discussed.

Based on the foregoing, this study proposes a flexible parametric Bayesian endogenous TQR model. 
The $p$-th quantile regression of interest is modelled parametrically following the usual Bayesian quantile regression approach. 
Following Lee~(2007), we introduce a control variable such that the conditional quantile of the error term is corrected to be zero and the parameters are correctly estimated. 
As in the approach of Lee~(2007), the $\alpha$-th quantile of the error term in the regression of the endogenous variable on the exogenous variables, which is often called the first stage regression, is also assumed to be zero.

We discuss the modelling approach for the first stage regression and consider a number of parametric and semiparametric models based on the extensions of Ogasawara and Kobayashi~(2015). 
Specifically, following  Wichitaksorn~\etal~(2014) and Naranjo~\etal~(2015), we employ the first stage regression models based on the asymmetric Laplace distribution, skew normal distribution, and asymmetric exponential power distribution, for which the $\alpha$-th quantile is always zero and is modelled by the regression function.  
To introduce more flexibility into the tail behaviour of the models based on the asymmetric Laplace and skew normal distributions, we also consider a semiparametric extension using the Dirichlet process mixture of scale parameters as in Kottas and Krnjaji\'c~(2011). 
The value of $\alpha$ is a priori unknown, while the choice of $\alpha$ can affect the estimates. 
In this study, hence, $\alpha$ is treated as a parameter to incorporate uncertainty and is estimated from the data. 
The performance of the proposed models is demonstrated in a simulation study under various settings, which is a novel contribution of the present study. 
We also illustrate the influence of the prior distributions on the posterior in the cases where valid and weak instruments are used. 

The rest of this paper is organised as follows. 
Section~\ref{sec:tobit} introduces the standard Bayesian TQR model with a motivating example. 
Then, Section~\ref{sec:approach} proposes Bayesian TQR models to deal with the endogenous variables. 
The MCMC methods adopted to make inferences about the models are also described. 
The simulation study under various settings is presented in Section~\ref{sec:sim}. 
The models are also illustrated by using the real data on the working hours of married women in Section~\ref{sec:real}. 
Finally, we conclude in Section~\ref{sec:conc}.  

\section{Bayesian TQR}\label{sec:tobit}
Suppose that the response variables are observed according to 
\begin{equation*}
y_i=c(y_i^*)=\max\left\{0,y_i^*\right\}, \quad i=1,\dots,n.
\end{equation*}
Then, consider the $p$-th quantile regression model for $y_i^*$ given by
\begin{equation*}
y_i^*=\vx_i'\vbeta_p+\eps_i, \quad i=1,\dots,n, 
\end{equation*}
where $\vx_i$ is the vector of regressors, $\vbeta_p$ is the coefficient parameter, and $\eps_i$ is the error term whose $p$-th quantile is zero. 
The $p$-th conditional quantile of $y^*$ is modelled as $Q_{y^*|\vx}(p)=\vx'\vbeta_p$. 
The equivariance under the monotone transformation $c(\cdot)$ of quantiles implies that the $p$-th conditional quantile of $y$ is given by
\begin{equation*}
Q_{y|\vx}(p)=c(Q_{y^*|\vx}(p)).
\end{equation*}
The TQR model can be estimated by minimising the sum of asymmetrically weighted absolute errors 
\begin{equation}\label{eqn:check}
\min_{\vbeta_p}\sum_{i=1}^n \rho_p(y_i-c(\vx_i'\vbeta_p)), 
\end{equation}
where $\rho_p(u)=u(p-I(u<0))$ and $I(\cdot)$ denotes the indicator function (Powell,~1986).

The Bayesian approach  assumes that $\eps$ follows the asymmetric Laplace distribution, since minimising (\ref{eqn:check}) is equivalent to maximising the likelihood function of the asymmetric Laplace distribution  (Koenker and Machado,~1999; Chernozhukov and Hong,~2003). 
The probability density function of the asymmetric Laplace distribution, denoted by $\AL(\sigma,p)$, is given by
\begin{equation}\label{eqn:ald}
f_{AL}(\eps|\sigma,p)=\frac{p(1-p)}{\sigma}\exp\left\{-\frac{\rho_p(\eps)}{\sigma}\right\}, \quad -\infty<x<\infty,
\end{equation}
where $\sigma>0$ is the scale parameter and $p\in(0,1)$ is the shape parameter (Yu and Zhang,~2005). 
The mean and variance are given by $E[\eps]=\sigma\frac{1-2p}{p(1-p)}$ and $\Var(\eps)=\sigma^2\frac{1-2p+2p^2}{p^2(1-p)^2}$.
The $p$-th quantile of this distribution is zero, $\int_{-\infty}^0f(\eps)=p$.
Assuming the prior distributions for the parameters, the parameters are estimated by using the MCMC method (\eg Yu and Stander,~2007; Kozumi and Kobayashi,~2011). 
Posterior consistency of Bayesian quantile regression based on the asymmetric Laplace distribution was shown by Sriram~\etal~(2013).

Estimates under the standard Bayesian TQR model are biased when  endogenous variables are included  as regressors. 
Consider a simple motivating example where the dataset was generated from 
\begin{equation}\label{eqn:example}
\begin{split}
y_i^*&=\beta_0+\beta_1x_i+\delta d_i+u_i, \\
d_i&=\gamma_0+\gamma_1x_i+\gamma_2w_i+v_i,\\
\end{split}
\end{equation}
for $i=1,\dots,300$, where $(\beta_0,\beta_1,\delta)=(1,1,1)$, $(\gamma_0,\gamma_1,\gamma_2)=(1,1,1)$, $x_i, w_i\sim \N(0,1)$ and 
\begin{equation*}
\left(
\begin{array}{c}
u_i\\v_i
\end{array}
\right)\sim\N(\zero,\mSigma), \quad \mSigma=\left[
\begin{array}{cc}
1&\rho \\
\rho&1
\end{array}
\right]. 
\end{equation*}
See also Chernozhukov~\etal~(2014). 
Note that $\rho$ expresses the level of endogeneity. 
While $d$ is an exogenous variable when $\rho=0$, $d$ is endogenous when $\rho\neq0$. 
Since $u|v\sim\N(\rho v,1-\rho^2)$, the model can be rewritten as 
\begin{equation}\label{eqn:sim1}
y_i^*=\beta_0+\beta_1x_i+\delta d_i+\rho v+\sqrt{1-\rho^2}u_i. 
\end{equation}
Therefore, the standard model that models the conditional quantile of $y^*$ as $\beta_0+\beta_1x+\delta d$ produces biased estimates. 

\begin{figure}[H]
\centering
\includegraphics[width=\textwidth]{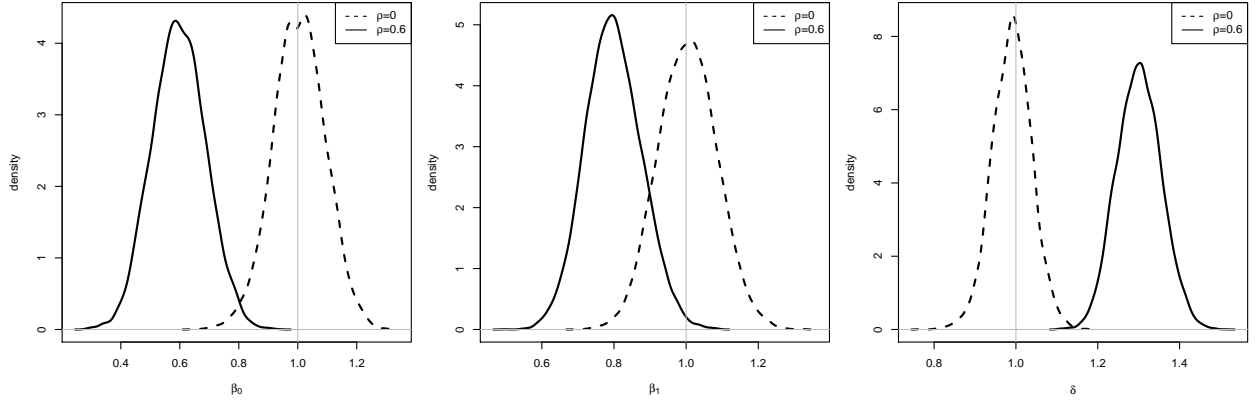}
\caption{Posterior distributions of $\beta_0$, $\beta_1$, and $\delta$ using the standard Bayesian Tobit median regression}\label{fig:fig1}
\end{figure}

Figure~\ref{fig:fig1} shows the posterior distributions of $\beta_0$, $\beta_1$, and $\delta$ for the standard model for $p=0.5$ obtained by using  the method of Kozumi and Kobayashi~(2011). 
The vertical lines in the figure indicate the true values.  
In the case of $\rho=0$, the posterior distributions are concentrated around the true values. 
However, in the case of $\rho=0.6$, the posterior distributions are concentrated away from the true values. 

\section{Bayesian Endogenous TQR Model}\label{sec:approach}
\subsection{Model}\label{sec:model}
We propose the following model to deal with the endogenous variables: 
\begin{eqnarray}
y_i^*&=&\vx_i'\vbeta_p+\delta_pd_i+\eta_p(d_i-\vz_i'\vgamma)+e_i, \label{eqn:2nd}\\
d_i&=&\vz_i'\vgamma+ v_i, \label{eqn:1st}
\end{eqnarray}
for $i=1,\dots,n$, where $\vx_i$ is the vector of the exogenous variables whose the first element is $1$, $d_i$ is the endogenous variable, $\vz_i=(\vx_i',w_i)'$, and $w_i$ is the exogenous variable not included in $\vx_i$, which is also called the instrumental variable. 
The term $d_i-\vz_i'\vgamma=v_i$ in (\ref{eqn:2nd}) is called the control variable and is introduced to account for endogeneity. 
Note that $\eta_p\neq0$ indicates $d_i$ is endogenous. 
We refer to (\ref{eqn:1st}) as the first stage regression and to (\ref{eqn:2nd}) as the second stage regression. 
A similar form is found in Lopes and Polson~(2014) in the context of the instrumental variable regression for means by using the Cholesky-based prior. 

Following Lee~(2007), the error term $\eps_i$ of the standard Bayesian TQR is decomposed into the terms $\eta_p(d_i-\vz_i'\vgamma)$ and $e_i$. 
It is assumed that relationship (\ref{eqn:1st}) is specified correctly and the quantile independence of $e_i$ on $\vz_i$ conditional on $v_i$: 
\begin{equation}
Q_{\eps|d,\vz}(p)=Q_{\eps|v,\vz}(p)=Q_{\eps|v}(p)=\eta_p(d-\vz'\vgamma).
\end{equation}
As in Lee~(2007), we also assume 
\begin{equation}\label{eqn:alpha}
Q_{v|\vz}(\alpha)=0, 
\end{equation}
where the $\alpha$-th conditional quantile of $v_i$ is zero for some $\alpha\in(0,1)$. 

\subsection{First Stage Regression}\label{sec:1st}
We are mainly concerned with modelling the first stage error that satisfies (\ref{eqn:alpha}). 
A simple and convenient approach is to assume $v_i\sim\AL(\phi,\alpha), \ i=1,\dots,n,$ as in Ogasawara and Kobayashi~(2015), since (\ref{eqn:alpha}) is always satisfied for the asymmetric Laplace distribution. 
However, the asymmetric Laplace distribution has limitations, such as peaky density, restrictive tail behaviour, and skewness. 
When a model lacks fit to the data, the estimate of the conditional quantile would be away from the value such that (\ref{eqn:alpha}) truly holds. 
Then, assuming $v_i$ is homoskedastic, the estimate of the intercept, $\gamma_0$, may be biased as well. 
Consequently, the estimate of $\vbeta_{p0}$ would be affected through the introduced term $\eta_p(d_i-\vz_i'\vgamma)$. 
When $v_i$ is heteroskedastic, the entire coefficient vector would be affected. 
Therefore, we consider some alternative models for the first stage error distribution.

Recently, Wichitaksorn~\etal~(2014) considered a class of parametric distributions with a quantile constraint of the form (\ref{eqn:alpha}), including the asymmetric Laplace distribution, and applied them in the context of quantile modelling. 
Furthermore, Zhu and Zinde-Walsh~(2009), Zhu and Galbraith~(2011), and Naranjo~\etal~(2015) considered a flexible parametric distribution with the quantile constraint. 
Based on these studies, we also consider the following two distributions to model the first stage error. 

First, we consider the skew normal distribution denoted by $\SN(\phi,\alpha)$, where $\phi>0$ is the scale parameter and $\alpha\in(0,1)$ is the shape parameter.
The probability density function is given by
\begin{equation}\label{eqn:sn}
f_{SN}(v|\phi,\alpha)=\frac{4\alpha(1-\alpha)}{\sqrt{2\pi\phi}}\exp\left\{-\frac{v^2}{2\phi}4(\alpha-I(v\leq0))^2\right\}. 
\end{equation}
When $\alpha=0.5$, the distribution reduces to $\N(0,\phi)$. 
The mean and variance are given by $E[v]=\sqrt{\frac{\phi}{2\pi}}\frac{1-2\alpha}{\alpha(1-\alpha)}$ and $\Var(v)=\phi\frac{\pi(1-3\alpha+3\alpha^2)-2(1-2\alpha)^2}{4\pi\alpha^2(1-\alpha)^2}$ (see Wichitaksorn~\etal,~2014). 
When the actual error distribution is close to the normal distribution, this distribution would lead to better performance than the asymmetric Laplace distribution. 
However, just as the asymmetric Laplace distribution, the skewness and the quantile level of the mode are controlled by the single parameter $\alpha$. 

Second, we consider the asymmetric exponential power distribution treated by Zhu and Zinde-Walsh~(2009), Zhu and Galbraith~(2011), and Naranjo~\etal~(2015). 
The probability density function of the asymmetric exponential power distribution, denoted by $\AEP(\phi,\alpha,\zeta_1,\zeta_2)$, is given by
\begin{equation}\label{eqn:aep}
f_{AEP}(v|\phi,\alpha,\zeta_1,\zeta_2)=\left\{
\begin{array}{ll}
\frac{1}{\phi}\exp\left\{-\left|\frac{v}{\alpha\phi/\Gamma(1+1/\zeta_1)}\right|^{\zeta_1}\right\},& \text{if}\quad v\leq0, \\
\frac{1}{\phi}\exp\left\{-\left|\frac{v}{(1-\alpha)\phi/\Gamma(1+1/\zeta_2)}\right|^{\zeta_2}\right\},& \text{if}\quad v>0, \\
\end{array}
\right.
\end{equation}
where $\phi>0$ is the scale parameter, $\alpha\in(0,1)$ is the skewness parameter, $\zeta_1>0$ is the shape parameter for the left tail, and $\zeta_2>0$ is the shape parameter for the right tail. 
After some reparameterisation, the distribution reduces to the asymmetric Laplace distribution when $\zeta_1=\zeta_2=1$ and to the skew normal distribution when $\zeta_1=\zeta_2=2$. 
The tails of the asymmetric exponential power distribution are controlled separately by $\zeta_1$ and $\zeta_2$, respectively, and the overall skewness is controlled by $\alpha$. 
Although the distribution is more flexible than the above two distributions, the posterior computation using MCMC would be inefficient, because it includes two additional shape parameters and it has no convenient mixture representation, apart from the mixture of uniforms that is inefficient, to facilitate an efficient MCMC algorithm. 
The computational efficiency is also compared in Section~\ref{sec:sim}.

In addition to the three parametric models, we also consider the semiparametric extension of the models based on the asymmetric Laplace and skew normal distributions to achieve both flexibility and computational efficiency. 
More specifically, the following two models using the Dirichlet process mixtures of scales are considered: 
\begin{eqnarray}\label{eqn:aldp}
&f_{ALDP}(v|G)=\int f_{AL}(v|\phi,\alpha) dG(\phi),\quad &G\sim\DP(a,G_0), \\
\label{eqn:sndp}
&f_{SNDP}(v|G)=\int f_{SN}(v|\phi,\alpha) dG(\phi),\quad &G\sim\DP(a,G_0), 
\end{eqnarray}
where $\DP(a,G_0)$ denotes the Dirichlet process with the precision parameter $a>0$ and the base measure $G_0$. 
For both models, we set $G_0=\IG(c_0,d_0)$ as it is computationally convenient. 
While those mixture models have the same limitation as the parametric versions in terms of skewness, they extend the tail behaviour of the error distribution preserving (\ref{eqn:alpha}) (Kottas and Krnjaji\'c,~2009). 
Hereafter, the models with the asymmetric Laplace, skew normal, and asymmetric exponential power first stage errors are respectively denoted by AL, SN, and AEP, and those with the Dirichlet process mixtures are denoted by ALDP and SNDP. 

We must take care when selecting the $\alpha$ value in (\ref{eqn:alpha}), as it is a part of the model specification and can thus affect the estimates (Lee,~2007). 
We treat $\alpha$ as a parameter and estimate its value along with the other parameters. 
Since $\alpha$ determines the quantile level of the mode for all models considered here, our approach to modelling the first stage regression can also be regarded as a kind of mode regression (see Wichitaksorn~\etal,~2014).

To gain further flexibility, we might extend the model through a fully nonparametric mixture. 
Several semiparametric models in the context of Bayesian quantile regression with exogenous variables have been proposed by Kottas and Gelfand~(2001), Kottas and Krnjaji\'c~(2009), and Reich~\etal~(2010). 
For example, Kottas and Krnjaji\'c~(2009) considered  the nonparametric mixture of uniform distributions for any unimodal density on the real line with the quantile restriction at the mode using the Dirichlet process mixture  (see also Kottas and Gelfand,~2001). 
In the more flexible model proposed by Reich~\etal~(2010), the mode of the error distribution does not have to coincide with zero. 
This is achieved by using a nonparametric mixture of the quantile-restricted two-component mixtures of normal distributions. 
However, their approaches are not directly applicable in the present context where the value of $\alpha$ is estimated. 
If we were to estimate the quantile level for which the quantile restriction holds, the computation under the former model is expected to be extremely inefficient and unstable as the model involves many indicator functions, and $\alpha$ and the intercept would be highly correlated. 
The intercept would not be identifiable in the latter model.

We could further extend the model to account for heteroskedasticity such that
\begin{equation}\label{eqn:hetero}
d_i=\vz_i'\vgamma+ \vz_i'\vkappa v_i, 
\end{equation}
for $i=1,\dots,n$, where $\vz_i'\vkappa>0$ for all $i$ and the first element of $\vkappa$ is fixed to one (\eg Reich,~2010).  
In this case, the $\alpha$-th quantile of $d$ is given by $Q_{d|\vz}(\alpha)=\vz_i'\vgamma+\vz_i'\vkappa Q_{v|\vz}(\alpha)=\vz_i'(\vgamma+\vkappa Q_{v|\vz}(\alpha))$ as in the usual quantile regression. 
However, since the first stage regression model is built based on (\ref{eqn:alpha}), models (\ref{eqn:1st}) and (\ref{eqn:hetero}) would produce identical estimates.

\subsection{Second Stage Regression}
We next turn to the model of the new second stage error, $e_i$, in (\ref{eqn:2nd}). 
Since the $p$-th conditional quantile of $e_i$ is now zero, we  assume that $e_i\sim\AL(\sigma,p), \ i=1,\dots,n,$  as in the standard Bayesian quantile regression approach. 
We utilise the location scale mixture of normals representation for the asymmetric Laplace distribution to facilitate an efficient MCMC method following Kozumi and Kobayashi~(2011) (see also Kotz~\etal,~2001). 
The model is expressed in the hierarchical form given by
\begin{eqnarray*}
y_i&=&\max\left\{y_i^*,0\right\},\\
y_i^*&\sim&\N(\tilde{\vx}_i'\tilde{\vbeta}_p+\theta_pg_i, \tau_p^2\sigma g_i),\\
g_i&\sim& \E(\sigma),
\end{eqnarray*}
for $i=1,\dots,n$, where $\tilde{\vx}_i=(\vx_i',d_i,d_i-\vz_i'\vgamma)'$, $\tilde{\vbeta}_p=(\vbeta_p',\delta_p,\eta_p)'$, $\E(\sigma)$ denotes the exponential distribution with mean $\sigma$, and 
\begin{equation}\label{eqn:theta}
\theta_u=\frac{1-2p}{p(1-p)},\quad \tau^2_p=\frac{2}{p(1-p)}. 
\end{equation}

\subsection{Prior Distributions}\label{sec:prior}
The coefficient parameter $\vgamma$ is common to all first stage regression specifications. 
First, we assume the normal prior for $\vgamma$, since it is computationally convenient for the AL, SN, ALDP, and SNDP models. 
Since we do not have information on the coefficient values, the variances are set such that the prior distributions are relatively diffuse. 
Our default choice is $\vgamma\sim\N(\zero,100\mI)$. 
For the scale parameters, $\phi$ for the AL, SN, and AEP distributions, a relatively diffuse inverse gamma distribution is assumed and the default choice is set to $\IG(0.1,0.1)$. 
For AEP, we assume $\zeta_j\sim\TN_{(0,\infty)}(1,1)$, where $\TN_{(a,b)}(\mu,\sigma^2)$ denotes the normal distribution with the mean $\mu$ and variance $\sigma^2$ truncated on the interval $(a,b)$. 
A similar prior specification is found in Naranjo~\etal~(2015). 
For all models, $\alpha\sim\U(0,1)$ is assumed. 

For the semiparametric models, we need to specify the parameters of the inverse gamma base measure. 
Assuming that the data have been rescaled, $c_0$ and $d_0$ are chosen such that the variance of $v_i$ takes values between $0$ and $3$ with high probability (\eg Ishwaran and James,~2002). 
Our default choice is $c_0=2$ and $d_0=0.5$ for ALDP and $c_0=d_0=1.5$ for SNDP. 
Under this choice, when $\alpha=0.5$ for ALDP, $\Pr(\phi_l\leq \sqrt{3.0/8})=0.802$ as $\Var(v_i)=8\phi^2$. 
Similarly, when $\alpha=0.4$, $\Pr(\phi_l\leq\sqrt{3.0/0.332})=0.784$. 
For SNDP, $\Pr(\phi_l\leq3)=0.801$ when $\alpha=0.5$ and $\Pr(\phi_l\leq 3/1.104)=0.775$ when $\alpha=0.4$. 
For the precision parameter of the Dirichlet process, $a$, we assume $a\sim\G(2,2)$ such that both small and large values for $a$, hence the number of clusters, are allowed.

For the coefficient parameters in the second stage, $\vbeta_p$ and $\delta_p$, we also assume relatively diffuse normal distributions. 
Our default choice of prior is $(\vbeta_p',\eta_p)'\sim\N(\zero,100\mI)$. 
Similar to $\phi$ in the parametric first stage, we assume an inverse gamma prior for the scale of the AL pseudo likelihood. 
Our default choice is $\IG(0.1,0.1)$. 

The parameter $\eta_p$ accounts for the endogeneity and we need to take care in prior elicitation. 
When the data follow the bivariate normal distribution, as in the motivating example (\ref{eqn:example}), $\eta_p$ is equal to $\rho\sigma_1/\sigma_2$, where $\rho$ is the correlation coefficient and $\sigma_1$ and $\sigma_2$ are the standard deviations of the first and second stage errors, respectively. 
In this case, we may follow Lopes and Polson~(2014) to determine the variance of the normal prior implied from an inverse Wishart prior for the covariance matrix. 
However, we do not limit ourselves to normal data as the quantile regression approach is suitable for heteroskedastic and non-normal data, and the non-normal models are used in the first stage. 
In the literature on Bayesian non-normal selection models, the prior distribution of $\eta_p$ is normal typically with a very small variance, such as $1/2$ (\eg Munkin and Trivedi,~2003, 2008; Deb~\etal,~2006).  
On the other hand, we use a more diffused prior to reflect our ignorance about $\eta_p$ and set our default choice of prior to be $\eta_p\sim\N(0,5)$. 
When the instrument is weak, it is expected that our quantile regression models face the problem of prior sensitivity and that the posterior distributions exhibit sharp behaviour, as in the case of the Bayesian instrumental variable regression model. 
Section~\ref{sec:sim} considers the alternative choices of the hyperparameters to study the prior sensitivity.

\subsection{MCMC Method}\label{sec:mcmc}
The proposed models are estimated by using the MCMC method based on the Gibbs sampler. 
We describe the Gibbs sampler for the semiparametric models with ALDP and SNDP, which is an extension of the Gibbs sampler described in Kozumi and Kobayashi~(2011) and Ogasawara and Kobayashi~(2015). 
The algorithms for the AL and SN models can be obtained straightforwardly. 
We also mention the algorithm for the AEP model. 

The variables involved in the Dirichlet process are sampled by using the retrospective sampler (Papaspiliopoulos and Roberts,~2008) and the slice sampler (Walker,~2007). 
First, we introduce $u_i\sim\U(0,1)$ and $k_i, \ i=1,\dots,n$, such that $\pi_l=\Pr(k_i=l), l=1,\dots,\infty$. 
Then, as in Walker~(2007), the Gibbs sampler is constructed by working on the following joint densities
\begin{eqnarray*}
f_{ALDP}(v_i,u_i)=\sum_{l=1}^\infty I(u_i<\omega_l) f_{AL}(v_i|\phi_l,\alpha),\\
f_{SNDP}(v_i,u_i)=\sum_{l=1}^\infty I(u_i<\omega_l) f_{SN}(v_i|\phi_l,\alpha),
\end{eqnarray*}
where $\phi_l\sim G_0$, $\pi_l=\omega_l\prod_{l<r}(1-\omega_r) $, $\omega_l\sim\B(1,a)$, and $\B(a,b)$ denotes the beta distribution with the parameters $a$ and $b$ (Sethuraman,~1994).  
We also let $k^*$ denote the minimum integer such that $\sum_{l=1}^{k^*}\pi_l>1-\min\left\{u_1,\dots,u_n\right\}$.

\subsubsection*{Algorithm for ALDP}
For the ALDP model, we utilise the mixture representation for the asymmetric Laplace distribution to sample $\vgamma$ efficiently such that $v_i|h_i\sim\N(\theta_\alpha h_i,\tau^2_a\phi_ih_i)$, $h_i\sim \E(\phi_i)$, $i=1,\dots,n$, where $\theta_a$ and $\tau^2_a$ are defined as in (\ref{eqn:theta}). 
Let us denote $\tilde{\vbeta}_p=(\vbeta_p',\delta_p,\eta_p)'$ and $\tilde{\vx}_i=(\vx_i',d_i,v_i-\vz_i'\vgamma)'$. 
Our Gibbs sampler proceeds by alternately sampling $\left\{u_i\right\}_{i=1}^n$, $\left\{\omega_l\right\}_{l=1}^{k^*}$, $\left\{k_i\right\}_{i=1}^n$, $\left\{\phi_l\right\}_{l=1}^{k^*}$, $a$,  $\vgamma$, $\left\{h_i\right\}_{i=1}^n$, $\alpha$,  $\left\{y_i^*\right\}_{i=1}^n$, $\tilde{\vbeta}_p$, $\sigma$, and $\left\{g_i\right\}_{i=1}^n$. 

\begin{itemize}
\item\textbf{Sampling $\left\{u_i\right\}_{i=1}^n$:} 
Generate $u_i$ from $\U(0,\pi_{k_i})$ for $i=1,\dots,n$. 

\item\textbf{Sampling $\left\{\omega_l\right\}_{l=1}^{k^*}$:}
Generate $\omega_l$ from $\B(1+n_l,n-\sum_{r\leq l}n_r+a)$  where $n_l=\sum_{i=1}^nI(k_i=l)$ for $l=1,\dots,k^*$.  

\item\textbf{Sampling $\left\{k_i\right\}_{i=1}^n$:}
Generate $k_i$ from the multinomial distribution with probabilities 
\begin{equation*}
\Pr(k_i=l)\propto f_{AL}(d_i-\vz_i'\vgamma|\phi_l,\alpha) I(u_i<\pi_l), \ l=1,\dots, k^*. 
\end{equation*}
for $i=1,\dots,n$. 

\item\textbf{Sampling $\left\{\phi_l\right\}_{l=1}^{k^*}$: }
Generate $\phi_l$ from $\IG(c_l,d_l)$ where
\begin{equation*}
c_l = 1.5n_l+c_0, \quad d_l=\sum_{i:k_i=l}\left[h_i+\frac{(d_i-\vz_i'\vgamma-\theta_\alpha h_i)^2}{2\tau_\alpha^2 h_i}\right] + d_0.
\end{equation*}

\item\textbf{Sampling $a$:}
Assuming the gamma prior, $\G(a_0,b_0)$, we use the method described by Escobar and West~(1995) to sample $a$. 
By introducing $c\sim\B(a+1,n)$, the full conditional distribution of $a$ is the mixture of two gamma distributions given by
\begin{equation*}
\varphi\G(a_0+n^*,b_0-\log c) + (1-\varphi)\G(a_0+n^*-1,b_0-\log c),
\end{equation*}
where $n^*$ is the number of distinct clusters and $\varphi/(1-\varphi)=(a_0+n^*-1)/(n(b_0-\log c))$.

\item\textbf{Sampling $\vgamma$:} 
Assuming $\vgamma\sim\N(\vg_0,\mG_0)$, $\vgamma$ is sampled from $\N(\vg_1,\mG_1)$ where
\begin{eqnarray*}
\mG_1&=&\left[\sum_{i=1}^n \vz_i\left( \frac{\eta_p^2}{\tau_p^2\sigma g_i}+\frac{1}{\tau^2_\alpha\phi_{k_i}h_i}\right)\vz_i' + \mG_0^{-1}\right]^{-1},\\
\vg_1&=&\mG_1\left[\sum_{i=1}^n\vz_i\left(-\frac{\eta_p(y_i^*-\vx_i'\vbeta_p - \eta_pd_i-\theta_pg_i)}{\tau_p^2\sigma g_i}+\frac{d_i-\theta_\alpha h_i}{\tau_\alpha^2\phi_{k_i}h_i} \right)+\mG_0^{-1}\vg_0\right], 
\end{eqnarray*}
as the density of the full conditional distribution denoted by $\pi(\vgamma|-)$ is given by
\begin{eqnarray*}
\pi(\vgamma|-)&\propto&\exp\left\{-\sum_{i=1}^n\frac{(y_i^*-\vx_i'\vbeta_p-\delta_pd_i-\eta_p(d_i-\vz_i'\vgamma)-\theta_pg_i)^2}{2\tau_p^2\sigma g_i}\right\}\\
&&\times\exp\left\{-\sum_{i=1}^n\frac{(d_i-\vz_i'\vgamma)^2}{2\tau_{\alpha}^2\phi_{k_i}h_i}\right\}\exp\left\{-\frac{1}{2}(\vgamma-\vg_0)'\mG_0^{-1}(\vgamma-\vg_0)\right\}\\
&\propto&\exp\left\{-\frac{1}{2}(\vgamma-\vg_1)'\mG_1^{-1}(\vgamma-\vg_1)\right\}. 
\end{eqnarray*}

\item\textbf{Sampling $\left\{h_i\right\}_{i=1}^n$:}
The full conditional distribution of $h_i$ is the generalised inverse Gaussian distribution, denoted by $\GIG(\nu,\xi,\chi)$. 
The probability density function of $\GIG(\nu, \xi,\chi)$ is given by
\begin{equation*}
f(x|\nu,\xi,\chi)=\frac{(\chi/\xi)^\nu}{2K_\nu(\xi\chi)}x^{\nu-1}\exp\left\{-\frac{1}{2}(\xi^2x^{-1}+\chi^2x)\right\},\quad x>0, \quad -\infty<\nu<\infty, \quad\xi,\chi\geq0,
\end{equation*}
where $K_\nu(\cdot)$ is the modified Bessel function of the third kind (Barndorff-Nielsen and Shephard,~2001). 
For $i=1,\dots,n$, we sample $h_i$ from $\GIG(1/2,\xi_i,\chi_i)$ where
\begin{equation*}
\xi_i^2=\frac{(d_i-\vz_i'\vgamma)^2}{\tau_\alpha^2\phi_{k_i}}, \quad 
\chi^2_i=\frac{\theta_a^2}{\tau_\alpha^2\phi_{k_i}}+\frac{2}{\phi_{k_i}}. 
\end{equation*}

\item\textbf{Sampling $\alpha$:}
The density of the full conditional distribution of $\alpha$ is given by
\begin{equation*}
\pi(\alpha|-)\propto\pi(\alpha)\prod_{i=1}^n f_{AL}(d_i-\vz_i'\vgamma|\phi_{k_i},\alpha),
\end{equation*}
where $\pi(\alpha|-)$ and$\pi(\alpha)$ denote the full conditional and  prior density of $\alpha$, respectively. 
We use the random walk Metropolis--Hastings (MH) algorithm to sample from this distribution.

\item\textbf{Sampling $\left\{y_i^*\right\}_{i=1}^n$:}
The full conditional distribution of $y_i^*$ is given by
\begin{equation*}
y_iI(y_i>0)+\TN_{(-\infty,0)}(\tilde{\vx}_i'\tilde{\vbeta}_p+\theta_pg_i, \tau_p^2\sigma g_i)I(y_i=0), \quad i=1,\dots,n. 
\end{equation*}

\item\textbf{Sampling $\tilde{\vbeta}_p$:}
We sample $\tilde{\vbeta}_p=(\vbeta_p',\delta_p,\eta_p)'$ in one block. 
Assuming $\tilde{\vbeta}_p\sim\N(\tilde{\vb}_0,\tilde{\mB}_0)$, the full conditional distribution is given by $\N(\tilde{\vb}_1,\tilde{\mB}_1)$ where
\begin{equation*}
\tilde{\mB}_1=\left[\sum_{i=1}^n\frac{\tilde{\vx}_i\tilde{\vx}_i'}{\tau_p^2\sigma g_i}+\tilde{\mB}_0^{-1}\right]^{-1}, \quad
\tilde{\vb}_1=\tilde{\mB}_1\left[\sum_{i=1}^n\frac{\tilde{\vx}_i(y_i^*-\theta_pg_i)}{\tau_p^2\sigma g_i}+ \tilde{\mB}_0^{-1}\tilde{\vb}_0\right]. 
\end{equation*}

\item\textbf{Sampling $\sigma$:}
Assuming $\sigma\sim\IG(m_0,s_0)$, we sample $\sigma$ from $\IG(m_1, s_1)$ where $m_1=1.5n+m_0$ and $s_1=\sum_{i=1}^ng_i+\sum_{i=1}^n(y_i-\tilde{\vx}_i'\tilde{\vbeta}_p-\theta_pg_i)^2/2\tau_p^2g_i+s_0$.

\item\textbf{Sampling $\left\{g_i\right\}_{i=1}^n$:}
Similar to $h_i$, $g_i$ is sampled from $\GIG( 1/2, \lambda_i,\psi)$ where
\begin{equation*}
\lambda_i^2=\frac{(y_i^*-\tilde{\vx}_i'\tilde{\vbeta}_p)^2}{\tau_p^2\sigma}, \quad \psi^2=\frac{\theta_p^2}{\tau_p^2\sigma}+\frac{2}{\sigma}, \quad i=1,\dots,n.
\end{equation*}
\end{itemize}

\subsubsection*{Algorithm for SNDP}
The Gibbs sampler for SNDP consists of sampling $\left\{u_i\right\}_{i=1}^n$, $\left\{\omega_l\right\}_{l=1}^{k^*}$, $\left\{k_i\right\}_{i=1}^n$, $\left\{\phi_l\right\}_{l=1}^{k^*}$, $a$,  $\vgamma$, $\alpha$,  $\left\{y_i^*\right\}_{i=1}^n$, $\tilde{\vbeta}_p$, $\sigma$, and $\left\{g_i\right\}_{i=1}^n$. 
The sampling algorithms for $\left\{u_i\right\}_{i=1}^n$, $\left\{\omega_l\right\}_{l=1}^{k^*}$, $a$, $\left\{y_i^*\right\}_{i=1}^n$, $\tilde{\vbeta}_p$, $\sigma$, and $\left\{g_i\right\}_{i=1}^n$ remain the same as in the case of ALDP. 
The sampling scheme of $\left\{k_i\right\}_{i=1}^n$ and $\alpha$ can be obtained by replacing $f_{AL}(d_i-\vz_i'\vgamma|\phi_{k_i},\alpha)$ with  $f_{SN}(d_i-\vz_i'\vgamma|\phi_{k_i},\alpha)$. 

Similar to the case of ALDP, the density of the full conditional distribution is given by
\begin{equation*}
\pi(\vgamma|-)\propto\exp\left\{-\frac{1}{2}(\vgamma-\vg_1(\vgamma))'\mG_1(\vgamma)^{-1}(\vgamma-\vg_1(\vgamma))\right\},
\end{equation*}
where
\begin{eqnarray*}
\mG_1(\vgamma)&=&\left[\sum_{i=1}^n \vz_i \left(\frac{\eta_p^2}{\tau_p^2\sigma g_i} +\frac{4(\alpha-I(d_i\leq\vz_i'\vgamma))^2}{\phi_{k_i}}  \right)\vz_i'+\mG_0^{-1}\right]^{-1}, \\
\vg_1(\vgamma)&=&\mG_1(\vgamma)\left[\sum_{i=1}^n\vz_i \left( -\frac{\eta_p(y_i^*-\vx_i'\vbeta_p-\eta_pd_i-\theta_p g_i)}{\tau_p^2\sigma g_i}+\frac{4d_i(\alpha-I(d_i\leq\vz_i'\vgamma))^2}{\phi_{k_i}}\right) + \mG_0^{-1}\vg_0\right], 
\end{eqnarray*}
which is similar to the density of the normal distribution. 
Therefore, we sample $\vgamma$ by using the MH algorithm with the proposal distribution given by $\N(\vg_1(\vgamma),\mG_1(\vgamma))$.

\subsubsection*{Algorithm for AEP}
Since no convenient representation for the AEP distribution is available, the full conditional distributions of the parameters in the first stage regression, $\vgamma$, $\phi$, $\alpha$, $\zeta_1$, and $\zeta_2$, are not in the standard forms. 
Therefore, we employ the adaptive random walk MH algorithm. 
Although Naranjo~\etal~(2015) proposed the scale mixture of uniform representation for the AEP distribution, the algorithm based on this representation would be inefficient, because it consists of sampling from a series of distributions that are truncated on some intervals such that the mixture representation holds and such intervals move quite slowly as sampling proceeds (see also Kobayashi,~2015). 
Since the additional shape parameters in AEP free up the role of $\alpha$, $\alpha$ controls the overall skewness by allocating the weights on the left and right sides of the mode. 
Hence, the MCMC sample would exhibit relatively high correlation between $\alpha$ and $\gamma_0$.

\section{Simulation Study}\label{sec:sim}
The models considered in the previous section are demonstrated using simulated data. 
The aims of this section are (1) to compare the performance of the proposed models (Section~\ref{sec:default}), 
(2) to study the sensitivity to the prior settings, and
(3) to illustrate the behaviour of the posterior distribution when the instrument is weak (Section~\ref{sec:alt}). 

\subsection{Settings}
The data are generated from the model given by
\begin{equation}\label{eqn:sim}
\begin{split}
y_i^*&=\beta_{0}+\beta_{1}x_i+\delta d_i + \eta v_i + e_i,\\
d_i &= \gamma_0 + \gamma_1 x_i + \gamma_2 w_i + v_i, 
\end{split}
\end{equation}
for $i=1,\dots,300$, where $(\gamma_0,\gamma_1,\gamma_2)=(0,1,1.5)$ assuming that a valid instrument is available, $(\beta_0,\beta_1,\delta,\eta)=(0,1,1,0.6)$, $x_i\sim\N(0,1)$, and $w_i\sim\TN_{(0,\infty)}(1,1)$. 
The performance of the models is compared by considering the various settings for $v_i$, while the distributions of $e_i$ are kept relatively simple in order that the true values of the quantile regression coefficients are tractable. 
The following five settings are considered:\\
\textbf{Setting~1}
$v_i\sim\N(0,1)$, $e_i\sim\N(0,1-\eta^2)$, \\
\textbf{Setting~2}
$v_i\sim t_4$, $e_i\sim t_6$, \\
\textbf{Setting~3}
$v_i\sim\ST(-0.430,1,0.980,4)$, $e_i\sim t_6$,  \\
\textbf{Setting~4}
$v_i\sim\N(0,(1+0.5w_i)^2)$, $e_i\sim\N(0,1-\eta^2)$,  \\
\textbf{Setting~5}
$v_i\sim\ST(-0.430,(1+0.5w_i)^2,0.980,4)$, $e_i\sim t_6$,  \\
where $\ST(\mu,\sigma^2,\alpha,\nu)$ denotes the skew $t$ distribution with the location parameter $\mu$, scale parameter $\sigma^2$, skewness parameter $\alpha=\delta/\sqrt{1-\delta^2}$, $\delta\in(-1,1)$, and degree of freedom $\nu$ (see Azzalini and Capitanio,~2003; Fr$\ddot{\text{u}}$hwirth-Schnatter and Pyne,~2010), and we set $\delta=0.7$. 
In Setting~1, the error terms follow the bivariate normal distribution as in the motivating example in Section~\ref{sec:tobit}. 
Setting~2 considers the fat tailed first stage regression. 
Setting~3 considers a more difficult situation where the first stage error is fat tailed and skewed. 
Setting~4 replaces the first stage error of Setting~1 with the heteroskedastic error with respect to the instrument. 
Setting~5 is also a challenging situation where the first stage error is fat tailed, skewed, and heteroskedastic. 
In Settings~3 and 5, the location parameters of the first stage error distributions are set such that the mode of $v_i$ is zero and the quantile level of the mode is $0.435$. 
The average censoring rates for the settings are around $0.25$. 
For each setting, the data are replicated $100$ times.

\subsection{Results under the Default Priors}\label{sec:default}
We first estimated the proposed models under the default prior specifications (see Section~\ref{sec:prior}) for $p=0.1$ and $0.5$ by running the MCMC for $20000$ iterations and discarding the first $5000$ draws as the burn-in period. 
The standard Bayesian TQR model was also estimated. 
The bias and root mean squared error (RMSE) of the parameters were computed over the $100$ replications. 
To assess the efficiency of the MCMC algorithm, we also recorded the inefficiency factor, which was defined as a ratio of the numerical variance of the sample mean of the Markov chain to the variance of the independence draws (Chib,~2001). 

Table~\ref{tab:default} presents the biases, RMSEs, and median inefficiency factors for the parameters over the $100$ replications. 
First, we examined the inefficiency factors. 
Overall, our sampling algorithms appear to be efficient, especially for AL, SN, ALDP, and SNDP. 
The table shows that the inefficiency factors for AL, SN, ALDP, and SNDP are reasonably small for $\beta_{p1}$, $\delta_p$, $\eta_p$, $\gamma_1$, and $\gamma_2$. 
Since $\alpha$ and $\gamma_0$ determine the quantile level of the mode and location of the mode, respectively, the MCMC sample exhibits correlation between $\alpha$ and $\gamma_0$ and this results in higher inefficiency factors for them. 
Hence, the inefficiency factors for $\beta_{p0}$ tend to be higher than those for the other parameters. 
This pattern is more profound in the case of AEP where the inefficiency factors for $\alpha$, $\gamma_0$, and $\beta_{p0}$ are quite high. 
Since the additional shape parameters in AEP free up the role of $\alpha$, the MCMC sample exhibits higher correlation between $\alpha$ and $\gamma_0$. 
Furthermore, the inefficiency factors for the other parameters for AEP are also higher than those for the other endogenous models.

Next, we turn to the performance of the models. 
As expected, TQR produces biased estimates in all cases. 
The RMSEs for the proposed endogenous models are generally larger for $p=0.1$, which is below the censoring point, than for $p=0.5$. 
The AL and ALDP models result in similar performance. 
The AEP model shows the largest RMSEs for $\gamma_0$ and $\beta_{p0}$ among the proposed models for all cases. 
Combined with the high inefficiency factors for those parameters, the convergence of the MCMC algorithm for AEP may be difficult to ensure in the given simulation setting. 
This finding suggests a considerable practical limitation and, thus, AEP will not be considered henceforth. 
The same limitation applies to the potentially more flexible nonparametric models discussed in Section~\ref{sec:1st}. 

Table~\ref{tab:default} also shows that the estimation of the first stage regression can influence the second stage parameters. 
For example, in Setting~1, the RMSEs for $\gamma_0$ for SN and SNDP are smaller than those for AL and ALDP, as the true model is the normal and thus SN and SNDP produce smaller RMSEs for $\beta_{p0}$. 
Similarly, in Setting~4, the RMSEs for $\beta_{p0}$ for SN and SNDP are smaller than those for AL and ALDP. 
In addition, the heteroskedasticity in the first stage influences the performance of the slope parameters, resulting in slightly smaller RMSEs for $\beta_{p1}$ for SN and SNDP than for AL and ALDP. 
However, the performance of the SN model becomes worse when the first stage error is fat tailed, since the skew normal distribution cannot accommodate a fat tailed distribution. 
While the results in Setting~2 are somewhat comparable across the models, the table shows that SN results in larger biases and RMSEs in Setting~3 and, especially, Setting~5. 
In Setting~3, SN results in larger RMSEs for $\beta_{p0}$ than for AL, ALDP, and SNDP. 
In Setting~5, given the heteroskedasticity of the first stage, the biases and RMSEs for the intercept and slope parameters for SN are larger than those for AL, ALDP, and SNDP. 
On the other hand, compared with SN, the semiparametric SNDP model is able to cope with fat tailed errors and this produces results comparable with those for AL and ALDP. 

While the models result in reasonable overall performance, the results for Settings~3 and 5 also illustrate the limitation of our modelling approach to some extent. 
In Setting~3, the models exhibit some bias in $\beta_{p0}$ because of the lack of fit in the first stage. 
This lack of fit, which is represented by the bias for $\gamma_0$, is reflected in the bias for $\beta_{p0}$. 
The entire coefficient vector may be influenced by this lack of fit in the first stage in the presence of heteroskedasticity as in Setting~5. 
The lack of fit in the first stage is also indicated by the biases in $\alpha$. 
This finding implies that an inflexible first stage model can fail to estimate the true quantile such that (\ref{eqn:alpha}) holds and that choosing the value of $\alpha$ a priori could lead to biased estimates (see the discussion in Section~\ref{sec:1st}).

\renewcommand{\baselinestretch}{1.0}
\renewcommand{\arraystretch}{0.75}
\renewcommand{\tabcolsep}{0.5mm}
\begin{table}[H]
\caption{Biases, RMSEs, and inefficiency factors under the default priors}
\label{tab:default}
\centering
{\scriptsize
\begin{tabular}{cccrrrrrrrrrrrrrrrrrr}\toprule
\multicolumn{3}{c}{}& \multicolumn{3}{c}{TQR} & \multicolumn{3}{c}{AL} & \multicolumn{3}{c}{SN} & \multicolumn{3}{c}{AEP} & \multicolumn{3}{c}{ALDP} & \multicolumn{3}{c}{SNDP} \\
\cmidrule(lr){4-6}\cmidrule(lr){7-9}\cmidrule(lr){10-12}\cmidrule(lr){13-15}\cmidrule(lr){16-18}\cmidrule(lr){19-21}
Setting & $p$ & Parameter & \multicolumn{1}{c}{Bias} & \multicolumn{1}{c}{RMSE}& \multicolumn{1}{c}{IF} & \multicolumn{1}{c}{Bias} & \multicolumn{1}{c}{RMSE} & \multicolumn{1}{c}{IF} & \multicolumn{1}{c}{Bias} & \multicolumn{1}{c}{RMSE} & \multicolumn{1}{c}{IF}& \multicolumn{1}{c}{Bias} & \multicolumn{1}{c}{RMSE} & \multicolumn{1}{c}{IF}& \multicolumn{1}{c}{Bias} & \multicolumn{1}{c}{RMSE} & \multicolumn{1}{c}{IF}& \multicolumn{1}{c}{Bias} & \multicolumn{1}{c}{RMSE} & \multicolumn{1}{c}{IF}\\\hline
1 & 0.1 & $\beta_{p0}$ & -0.474 & 0.511 & 37.1 &  0.048 & 0.239 &  55.5 &  0.047 & 0.211 &  59.7 &  0.066 & 0.252 & 245.0 &  0.047 & 0.237 &  57.1 &  0.047 & 0.209 &  61.8 \\
  &     & $\beta_{p1}$ & -0.248 & 0.272 & 17.0 & -0.022 & 0.139 &  22.3 & -0.020 & 0.134 &  18.9 & -0.020 & 0.136 &  43.1 & -0.022 & 0.140 &  24.7 & -0.020 & 0.135 &  20.1 \\
  &     & $\delta_p$   &  0.200 & 0.212 & 27.4 & -0.009 & 0.092 &  24.9 & -0.007 & 0.085 &  20.0 & -0.007 & 0.085 &  46.3 & -0.008 & 0.092 &  24.6 & -0.007 & 0.085 &  21.1 \\
  &     & $\eta_p$     &        &       &      &  0.001 & 0.122 &  18.5 & -0.001 & 0.120 &  14.5 & -0.001 & 0.120 &  29.8 &  0.000 & 0.122 &  17.5 & -0.001 & 0.120 &  16.1 \\
  &     & $\gamma_0$   &        &       &      &  0.001 & 0.204 &  54.7 &  0.000 & 0.165 &  59.2 &  0.036 & 0.264 & 340.8 &  0.000 & 0.206 &  53.2 &  0.003 & 0.160 &  60.7 \\
  &     & $\gamma_1$   &        &       &      & -0.012 & 0.066 &  16.7 & -0.006 & 0.058 &   9.2 & -0.007 & 0.060 &  96.3 & -0.012 & 0.067 &  17.7 & -0.007 & 0.059 &   9.6 \\
  &     & $\gamma_2$   &        &       &      & -0.004 & 0.086 &  17.3 & -0.002 & 0.074 &   9.2 & -0.002 & 0.075 &  93.7 & -0.004 & 0.086 &  16.9 & -0.003 & 0.074 &   8.9 \\
  &     & $\alpha$     &        &       &      & -0.002 & 0.052 &  66.1 & -0.001 & 0.043 &  72.2 &  0.011 & 0.089 & 357.1 & -0.002 & 0.054 &  65.0 & -0.000 & 0.042 &  76.4 \\\cmidrule{2-21}
  & 0.5 & $\beta_{p0}$ & -0.426 & 0.443 & 11.7 &  0.017 & 0.180 &  25.5 &  0.018 & 0.167 &  34.6 &  0.030 & 0.196 & 243.7 &  0.017 & 0.182 &  25.3 &  0.017 & 0.165 &  34.1 \\
  &     & $\beta_{p1}$ & -0.235 & 0.251 &  7.9 & -0.001 & 0.089 &  12.8 &  0.001 & 0.087 &   9.9 &  0.001 & 0.088 &  27.3 & -0.001 & 0.089 &  12.1 &  0.001 & 0.086 &   9.3 \\
  &     & $\delta_p$   &  0.233 & 0.238 &  9.7 & -0.004 & 0.063 &  11.6 & -0.003 & 0.061 &  10.1 & -0.003 & 0.062 &  28.7 & -0.004 & 0.063 &  11.6 & -0.003 & 0.061 &   9.2 \\
  &     & $\eta_p$     &        &       &      &  0.004 & 0.086 &   8.7 &  0.003 & 0.084 &   8.2 &  0.003 & 0.085 &  18.3 &  0.004 & 0.086 &   8.2 &  0.003 & 0.083 &   7.6 \\
  &     & $\gamma_0$   &        &       &      &  0.003 & 0.206 &  37.6 &  0.003 & 0.163 &  44.2 &  0.028 & 0.254 & 313.0 &  0.003 & 0.209 &  41.0 &  0.003 & 0.161 &  47.4 \\
  &     & $\gamma_1$   &        &       &      & -0.012 & 0.066 &  13.1 & -0.006 & 0.058 &   5.7 & -0.008 & 0.060 &  74.6 & -0.012 & 0.066 &  12.2 & -0.007 & 0.059 &   5.8 \\
  &     & $\gamma_2$   &        &       &      & -0.005 & 0.085 &  12.0 & -0.003 & 0.074 &   5.2 & -0.003 & 0.075 &  60.5 & -0.005 & 0.086 &  13.5 & -0.003 & 0.074 &   5.9 \\
  &     & $\alpha$     &        &       &      & -0.001 & 0.053 &  53.7 & -0.000 & 0.043 &  57.1 &  0.008 & 0.086 & 328.5 & -0.001 & 0.055 &  50.4 & -0.000 & 0.042 &  62.3 \\\hline
2 & 0.1 & $\beta_{p0}$ & -0.594 & 0.657 & 53.6 &  0.088 & 0.302 &  40.5 &  0.082 & 0.309 &  50.2 &  0.099 & 0.354 & 120.4 &  0.090 & 0.304 &  40.6 &  0.096 & 0.305 &  47.6 \\
  &     & $\beta_{p1}$ & -0.297 & 0.341 & 18.7 & -0.009 & 0.161 &  22.3 & -0.009 & 0.164 &  21.6 & -0.008 & 0.159 &  38.0 & -0.008 & 0.160 &  20.0 & -0.010 & 0.160 &  21.1 \\
  &     & $\delta_p$   &  0.268 & 0.282 & 38.7 & -0.025 & 0.115 &  23.4 & -0.023 & 0.115 &  25.9 & -0.024 & 0.115 &  35.9 & -0.024 & 0.115 &  24.6 & -0.023 & 0.115 &  24.1 \\
  &     & $\eta_p$     &        &       &      &  0.005 & 0.139 &  19.0 &  0.003 & 0.138 &  18.9 &  0.005 & 0.138 &  27.2 &  0.005 & 0.138 &  19.7 &  0.004 & 0.137 &  19.8 \\
  &     & $\gamma_0$   &        &       &      & -0.025 & 0.189 &  27.8 & -0.038 & 0.244 &  25.8 & -0.001 & 0.339 & 203.2 & -0.022 & 0.191 &  25.7 & -0.015 & 0.176 &  32.6 \\
  &     & $\gamma_1$   &        &       &      &  0.001 & 0.073 &  11.6 &  0.003 & 0.082 &   7.8 &  0.000 & 0.070 &  70.1 &  0.001 & 0.073 &  12.4 & -0.001 & 0.070 &   8.4 \\
  &     & $\gamma_2$   &        &       &      & -0.002 & 0.092 &  12.1 &  0.004 & 0.094 &   8.2 & -0.001 & 0.089 &  61.2 & -0.002 & 0.092 &  14.6 &  0.002 & 0.087 &   8.1 \\
  &     & $\alpha$     &        &       &      & -0.004 & 0.041 &  32.3 & -0.005 & 0.059 &  32.3 &  0.004 & 0.107 & 212.1 & -0.003 & 0.041 &  34.2 & -0.000 & 0.036 &  46.2 \\\cmidrule{2-21}
  & 0.5 & $\beta_{p0}$ & -0.579 & 0.604 & 13.7 & -0.001 & 0.198 &  16.6 & -0.013 & 0.207 &  17.1 &  0.013 & 0.288 &  98.1 & -0.000 & 0.200 &  17.4 &  0.004 & 0.193 &  18.9 \\
  &     & $\beta_{p1}$ & -0.302 & 0.323 &  6.9 &  0.003 & 0.127 &   9.4 &  0.002 & 0.130 &   9.2 &  0.001 & 0.123 &  22.2 &  0.002 & 0.126 &   9.8 & -0.001 & 0.123 &   8.1 \\
  &     & $\delta_p$   &  0.312 & 0.319 & 11.3 & -0.004 & 0.082 &   9.0 & -0.001 & 0.083 &   8.6 & -0.002 & 0.080 &  20.3 & -0.003 & 0.082 &   9.7 & -0.001 & 0.080 &   8.4 \\
  &     & $\eta_p$     &        &       &      &  0.009 & 0.099 &   8.1 &  0.007 & 0.099 &   7.5 &  0.008 & 0.097 &  14.6 &  0.009 & 0.099 &   8.2 &  0.007 & 0.097 &   7.3 \\
  &     & $\gamma_0$   &        &       &      & -0.025 & 0.188 &  20.8 & -0.041 & 0.245 &  17.8 & -0.003 & 0.341 & 165.9 & -0.023 & 0.191 &  25.3 & -0.015 & 0.176 &  24.0 \\
  &     & $\gamma_1$   &        &       &      &  0.001 & 0.073 &   9.2 &  0.003 & 0.082 &   5.0 &  0.000 & 0.070 &  58.6 &  0.001 & 0.073 &  10.7 & -0.001 & 0.070 &   5.8 \\
  &     & $\gamma_2$   &        &       &      & -0.002 & 0.091 &   9.3 &  0.005 & 0.094 &   4.7 &  0.001 & 0.089 &  50.2 & -0.002 & 0.092 &  10.4 &  0.002 & 0.087 &   5.3 \\
  &     & $\alpha$     &        &       &      & -0.004 & 0.041 &  29.7 & -0.005 & 0.059 &  26.4 &  0.004 & 0.108 & 193.7 & -0.003 & 0.041 &  35.8 & -0.000 & 0.036 &  40.7 \\\hline
3 & 0.1 & $\beta_{p0}$ & -0.464 & 0.539 & 36.0 &  0.028 & 0.287 &  33.0 & -0.006 & 0.301 &  36.4 &  0.113 & 0.334 & 108.2 &  0.027 & 0.288 &  35.4 &  0.026 & 0.282 &  37.7 \\
  &     & $\beta_{p1}$ & -0.264 & 0.314 & 18.9 & -0.007 & 0.180 &  18.1 & -0.007 & 0.182 &  18.7 & -0.010 & 0.181 &  31.5 & -0.008 & 0.180 &  19.2 & -0.009 & 0.182 &  17.8 \\
  &     & $\delta_p$   &  0.235 & 0.253 & 26.8 & -0.022 & 0.120 &  17.4 & -0.022 & 0.118 &  18.0 & -0.021 & 0.118 &  27.9 & -0.021 & 0.120 &  20.1 & -0.021 & 0.118 &  18.7 \\
  &     & $\eta_p$     &        &       &      & -0.011 & 0.149 &  14.5 & -0.010 & 0.147 &  15.0 & -0.012 & 0.148 &  21.7 & -0.012 & 0.149 &  16.4 & -0.011 & 0.147 &  12.8 \\
  &     & $\gamma_0$   &        &       &      & -0.096 & 0.201 &  23.6 & -0.147 & 0.285 &  23.9 &  0.041 & 0.309 & 199.4 & -0.099 & 0.207 &  25.5 & -0.098 & 0.186 &  27.7 \\
  &     & $\gamma_1$   &        &       &      &  0.001 & 0.058 &  10.9 &  1.001 & 1.003 &   6.4 & -0.001 & 0.055 &  66.7 &  0.000 & 0.058 &  10.9 & -0.001 & 0.057 &   5.4 \\
  &     & $\gamma_2$   &        &       &      & -0.002 & 0.084 &   9.8 &  1.496 & 1.499 &   5.2 & -0.001 & 0.079 &  60.1 & -0.002 & 0.086 &  11.0 & -0.002 & 0.077 &   6.1 \\
  &     & $\alpha$     &        &       &      & -0.041 & 0.060 &  33.4 & -0.055 & 0.089 &  35.1 &  0.020 & 0.116 & 214.8 & -0.042 & 0.061 &  35.5 & -0.039 & 0.055 &  48.0 \\\cmidrule{2-21}
  & 0.5 & $\beta_{p0}$ & -0.491 & 0.520 & 10.0 & -0.053 & 0.191 &  13.6 & -0.084 & 0.223 &  13.9 &  0.027 & 0.255 &  85.8 & -0.054 & 0.192 &  14.5 & -0.054 & 0.183 &  17.3 \\
  &     & $\beta_{p1}$ & -0.268 & 0.288 &  6.7 &  0.014 & 0.117 &   8.6 &  0.015 & 0.123 &   7.3 &  0.011 & 0.117 &  18.7 &  0.013 & 0.117 &   9.3 &  0.012 & 0.120 &   6.6 \\
  &     & $\delta_p$   &  0.292 & 0.298 &  7.7 & -0.006 & 0.072 &   9.2 & -0.007 & 0.071 &   7.2 & -0.006 & 0.073 &  16.4 & -0.006 & 0.072 &   8.6 & -0.007 & 0.072 &   7.5 \\
  &     & $\eta_p$     &        &       &      &  0.009 & 0.101 &   7.3 &  0.010 & 0.100 &   6.2 &  0.009 & 0.101 &  11.9 &  0.009 & 0.101 &   6.9 &  0.009 & 0.101 &   5.9 \\
  &     & $\gamma_0$   &        &       &      & -0.096 & 0.202 &  19.1 & -0.144 & 0.284 &  15.7 &  0.033 & 0.315 & 174.7 & -0.098 & 0.206 &  24.6 & -0.098 & 0.185 &  24.8 \\
  &     & $\gamma_1$   &        &       &      &  0.001 & 0.058 &   8.2 &  1.001 & 1.003 &   3.6 & -0.002 & 0.055 &  51.5 &  0.001 & 0.058 &   9.9 & -0.001 & 0.057 &   4.6 \\
  &     & $\gamma_2$   &        &       &      & -0.002 & 0.084 &   8.7 &  1.496 & 1.499 &   3.9 & -0.001 & 0.078 &  47.4 & -0.002 & 0.086 &   9.1 & -0.002 & 0.077 &   5.0 \\
  &     & $\alpha$     &        &       &      & -0.040 & 0.060 &  27.7 & -0.055 & 0.089 &  26.7 &  0.017 & 0.118 & 180.6 & -0.041 & 0.060 &  34.9 & -0.039 & 0.055 &  44.5 \\\hline
4 & 0.1 & $\beta_{p0}$ & -0.888 & 0.908 & 50.5 &  0.054 & 0.246 &  61.1 &  0.064 & 0.232 &  91.1 &  0.058 & 0.325 & 440.6 &  0.054 & 0.246 &  64.5 &  0.062 & 0.239 &  88.6 \\
  &     & $\beta_{p1}$ & -0.401 & 0.416 & 20.1 &  0.001 & 0.148 &  49.7 &  0.006 & 0.143 &  53.1 &  0.007 & 0.144 & 114.2 &  0.002 & 0.153 &  42.7 &  0.005 & 0.144 &  52.9 \\
  &     & $\delta_p$   &  0.369 & 0.374 & 35.8 & -0.018 & 0.101 &  49.0 & -0.019 & 0.101 &  49.0 & -0.019 & 0.099 & 128.8 & -0.018 & 0.102 &  44.8 & -0.017 & 0.098 &  56.2 \\
  &     & $\eta_p$     &        &       &      &  0.017 & 0.117 &  47.2 &  0.018 & 0.116 &  52.0 &  0.018 & 0.114 & 113.2 &  0.016 & 0.117 &  46.7 &  0.016 & 0.113 &  54.3 \\
  &     & $\gamma_0$   &        &       &      & -0.005 & 0.249 &  59.0 &  0.010 & 0.217 &  97.9 & -0.008 & 0.412 & 528.2 & -0.005 & 0.248 &  58.2 &  0.009 & 0.227 &  92.4 \\
  &     & $\gamma_1$   &        &       &      & -0.005 & 0.105 &  32.5 & -0.001 & 0.096 &  27.1 & -0.000 & 0.096 & 146.3 & -0.005 & 0.106 &  31.5 & -0.001 & 0.093 &  26.3 \\
  &     & $\gamma_2$   &        &       &      &  0.016 & 0.188 &  52.3 &  0.009 & 0.167 &  55.1 &  0.012 & 0.170 & 181.4 &  0.017 & 0.191 &  51.5 &  0.014 & 0.169 &  56.7 \\
  &     & $\alpha$     &        &       &      &  0.001 & 0.060 & 104.4 &  0.003 & 0.048 & 166.2 & -0.001 & 0.090 & 549.6 &  0.002 & 0.061 & 113.3 &  0.004 & 0.053 & 148.9 \\\cmidrule{2-21}
  & 0.5 & $\beta_{p0}$ & -0.676 & 0.685 & 15.1 &  0.005 & 0.201 &  31.2 &  0.018 & 0.190 &  48.6 &  0.016 & 0.296 & 355.5 &  0.005 & 0.203 &  33.9 &  0.014 & 0.189 &  44.8 \\
  &     & $\beta_{p1}$ & -0.388 & 0.397 &  7.7 & -0.003 & 0.141 &  25.7 & -0.001 & 0.129 &  24.2 & -0.002 & 0.130 &  70.0 & -0.002 & 0.144 &  23.7 & -0.003 & 0.129 &  26.4 \\
  &     & $\delta_p$   &  0.380 & 0.382 & 11.7 & -0.009 & 0.096 &  29.7 & -0.010 & 0.088 &  28.4 & -0.008 & 0.084 &  75.9 & -0.010 & 0.098 &  28.6 & -0.008 & 0.085 &  33.0 \\
  &     & $\eta_p$     &        &       &      &  0.023 & 0.106 &  26.5 &  0.024 & 0.098 &  27.0 &  0.022 & 0.095 &  65.9 &  0.024 & 0.108 &  25.8 &  0.022 & 0.095 &  29.8 \\
  &     & $\gamma_0$   &        &       &      & -0.005 & 0.248 &  34.3 &  0.015 & 0.220 &  58.4 &  0.008 & 0.429 & 406.3 & -0.004 & 0.251 &  37.7 &  0.011 & 0.220 &  57.0 \\
  &     & $\gamma_1$   &        &       &      & -0.005 & 0.105 &  20.0 & -0.001 & 0.095 &  14.1 & -0.001 & 0.095 &  96.3 & -0.004 & 0.105 &  23.2 & -0.001 & 0.093 &  15.8 \\
  &     & $\gamma_2$   &        &       &      &  0.017 & 0.188 &  36.3 &  0.011 & 0.164 &  32.6 &  0.015 & 0.166 & 114.9 &  0.017 & 0.192 &  32.7 &  0.015 & 0.167 &  39.3 \\
  &     & $\alpha$     &        &       &      &  0.001 & 0.059 &  76.6 &  0.004 & 0.048 &  99.8 &  0.003 & 0.094 & 444.4 &  0.002 & 0.062 &  76.5 &  0.004 & 0.051 & 103.7 \\\hline
5 & 0.1 & $\beta_{p0}$ & -0.853 & 0.900 & 41.8 &  0.016 & 0.299 &  38.2 & -0.032 & 0.339 &  47.6 &  0.155 & 0.408 & 158.2 &  0.014 & 0.298 &  39.0 & -0.004 & 0.295 &  39.1 \\
  &     & $\beta_{p1}$ & -0.405 & 0.438 & 20.9 &  0.028 & 0.198 &  27.1 &  0.045 & 0.222 &  33.3 &  0.029 & 0.202 &  58.1 &  0.030 & 0.198 &  25.1 &  0.030 & 0.205 &  25.9 \\
  &     & $\delta_p$   &  0.393 & 0.401 & 30.3 & -0.048 & 0.146 &  26.6 & -0.066 & 0.161 &  28.5 & -0.049 & 0.145 &  61.1 & -0.050 & 0.146 &  27.7 & -0.051 & 0.148 &  26.4 \\
  &     & $\eta_p$     &        &       &      &  0.028 & 0.158 &  26.7 &  0.046 & 0.170 &  27.3 &  0.029 & 0.157 &  55.4 &  0.029 & 0.158 &  25.8 &  0.031 & 0.159 &  25.1 \\
  &     & $\gamma_0$   &        &       &      & -0.093 & 0.240 &  26.8 & -0.159 & 0.360 &  34.5 &  0.124 & 0.412 & 257.2 & -0.097 & 0.247 &  25.5 & -0.120 & 0.243 &  31.1 \\
  &     & $\gamma_1$   &        &       &      &  0.001 & 0.088 &  17.1 & -0.000 & 0.127 &  16.7 &  0.001 & 0.087 &  74.4 &  0.001 & 0.088 &  16.8 & -0.001 & 0.090 &  11.9 \\
  &     & $\gamma_2$   &        &       &      & -0.056 & 0.160 &  24.9 & -0.081 & 0.202 &  22.4 & -0.055 & 0.156 & 104.9 & -0.058 & 0.164 &  25.5 & -0.064 & 0.161 &  20.6 \\
  &     & $\alpha$     &        &       &      & -0.041 & 0.061 &  51.3 & -0.057 & 0.094 &  57.3 &  0.020 & 0.101 & 288.7 & -0.043 & 0.063 &  49.8 & -0.048 & 0.066 &  65.7 \\\cmidrule{2-21}
  & 0.5 & $\beta_{p0}$ & -0.754 & 0.768 &  9.8 & -0.042 & 0.201 &  16.0 & -0.088 & 0.257 &  21.8 &  0.093 & 0.333 & 123.5 & -0.043 & 0.206 &  17.3 & -0.061 & 0.200 &  21.5 \\
  &     & $\beta_{p1}$ & -0.394 & 0.406 &  7.0 &  0.052 & 0.148 &  15.2 &  0.068 & 0.181 &  13.9 &  0.049 & 0.147 &  35.1 &  0.052 & 0.148 &  14.4 &  0.052 & 0.154 &  12.1 \\
  &     & $\delta_p$   &  0.430 & 0.432 &  7.6 & -0.042 & 0.102 &  15.1 & -0.059 & 0.124 &  15.0 & -0.040 & 0.100 &  41.7 & -0.042 & 0.101 &  16.2 & -0.045 & 0.105 &  14.0 \\
  &     & $\eta_p$     &        &       &      &  0.042 & 0.115 &  13.5 &  0.059 & 0.134 &  15.2 &  0.040 & 0.112 &  33.1 &  0.042 & 0.115 &  15.3 &  0.045 & 0.117 &  13.2 \\
  &     & $\gamma_0$   &        &       &      & -0.093 & 0.239 &  18.7 & -0.158 & 0.361 &  19.9 &  0.113 & 0.420 & 177.0 & -0.095 & 0.247 &  21.4 & -0.122 & 0.243 &  26.9 \\
  &     & $\gamma_1$   &        &       &      &  0.002 & 0.088 &  12.6 & -0.000 & 0.127 &   9.5 &  0.000 & 0.086 &  59.9 &  0.001 & 0.088 &  14.4 & -0.002 & 0.090 &   8.8 \\
  &     & $\gamma_2$   &        &       &      & -0.058 & 0.162 &  19.3 & -0.081 & 0.203 &  14.7 & -0.052 & 0.155 &  75.0 & -0.058 & 0.163 &  21.9 & -0.064 & 0.161 &  16.3 \\
  &     & $\alpha$     &        &       &      & -0.042 & 0.062 &  38.0 & -0.057 & 0.094 &  42.0 &  0.018 & 0.104 & 209.2 & -0.043 & 0.063 &  44.4 & -0.049 & 0.067 &  54.2 \\\bottomrule
\end{tabular}
}
\end{table}

\subsection{Alternative Base Measures and Prior Specifications}\label{sec:alt}
For comparison purposes, we consider two alternative specifications for the inverse gamma base measure for the semiparametric models. 
The following slightly less diffuse settings than the default are considered. 
For ALDP, we consider $\IG(2.5,0.6)$ such that $\Pr(\phi_l\leq\sqrt{3/8})=0.854$ and  $\IG(3.0,0.7)$ such that $\Pr(\phi_l\leq\sqrt{3/8})=0.891$ when $\alpha=0.5$.
For SNDP, we consider $\IG(2,2)$ such that $\Pr(\phi\leq3)=0.852$ and $\IG(2.5,2.5)$ such that $\Pr(\phi_l\leq3)=0.893$. 
For the other parameters, we use the default prior specifications. 
Table~\ref{tab:base} presents the biases and RMSEs for ALDP and SNDP under the alternative base measures for $p=0.1$ and $p=0.5$. 
The results in Table~\ref{tab:base} are essentially identical to those in Table~\ref{tab:default}, suggesting that the default choice of the base measures provides reasonable performance.

\renewcommand{\baselinestretch}{1.0}
\renewcommand{\arraystretch}{1.0}
\renewcommand{\tabcolsep}{0.5mm}
\begin{table}[H]
\caption{Biases and RMSEs for ALDP and SNDP under the alternative base measures}
\label{tab:base}
\centering
{\scriptsize
\begin{tabular}{cccrrrrrrrr}\toprule
\multicolumn{3}{c}{}& \multicolumn{4}{c}{ALDP} & \multicolumn{4}{c}{SNDP}\\
\cmidrule(lr){4-7}\cmidrule(lr){8-11}
\multicolumn{3}{c}{}&\multicolumn{2}{c}{Alternative 1}&\multicolumn{2}{c}{Alternative 2}&\multicolumn{2}{c}{Alternative 1}&\multicolumn{2}{c}{Alternative 2}\\
\cmidrule(lr){4-5}\cmidrule(lr){6-7}\cmidrule(lr){8-9}\cmidrule(lr){10-11}
Setting & $p$ & Parameter & \multicolumn{1}{c}{Bias} & \multicolumn{1}{c}{RMSE} & \multicolumn{1}{c}{Bias} & \multicolumn{1}{c}{RMSE} & \multicolumn{1}{c}{Bias} & \multicolumn{1}{c}{RMSE} & \multicolumn{1}{c}{Bias} & \multicolumn{1}{c}{RMSE} \\\hline
1 & 0.1 & $\beta_{p0}$ &  0.050 & 0.239 &  0.048 & 0.239 &  0.050 & 0.211 &  0.047 & 0.209 \\
  &     & $\beta_{p1}$ & -0.022 & 0.140 & -0.022 & 0.139 & -0.020 & 0.136 & -0.020 & 0.135 \\
  &     & $\delta_p$   & -0.008 & 0.092 & -0.009 & 0.092 & -0.008 & 0.085 & -0.007 & 0.086 \\\cmidrule{2-11}
  & 0.5 & $\beta_{p0}$ &  0.018 & 0.182 &  0.017 & 0.183 &  0.017 & 0.164 &  0.019 & 0.164 \\
  &     & $\beta_{p1}$ & -0.002 & 0.089 & -0.001 & 0.089 &  0.001 & 0.087 &  0.001 & 0.087 \\
  &     & $\delta_p$   & -0.004 & 0.063 & -0.004 & 0.063 & -0.003 & 0.061 & -0.003 & 0.061 \\\hline
2 & 0.1 & $\beta_{p0}$ &  0.092 & 0.304 &  0.088 & 0.304 &  0.091 & 0.302 &  0.094 & 0.302 \\
  &     & $\beta_{p1}$ & -0.008 & 0.160 & -0.007 & 0.160 & -0.011 & 0.159 & -0.012 & 0.158 \\
  &     & $\delta_p$   & -0.025 & 0.115 & -0.025 & 0.116 & -0.022 & 0.114 & -0.023 & 0.114 \\\cmidrule{2-11}
  & 0.5 & $\beta_{p0}$ & -0.001 & 0.199 & -0.001 & 0.200 &  0.003 & 0.194 &  0.002 & 0.194 \\
  &     & $\beta_{p1}$ &  0.003 & 0.127 &  0.003 & 0.126 &  0.000 & 0.124 & -0.001 & 0.123 \\
  &     & $\delta_p$   & -0.003 & 0.082 & -0.004 & 0.082 & -0.002 & 0.081 & -0.001 & 0.081 \\\hline
3 & 0.1 & $\beta_{p0}$ &  0.025 & 0.286 &  0.025 & 0.287 &  0.024 & 0.280 &  0.022 & 0.282 \\
  &     & $\beta_{p1}$ & -0.008 & 0.179 & -0.009 & 0.180 & -0.010 & 0.182 & -0.010 & 0.182 \\
  &     & $\delta_p$   & -0.020 & 0.119 & -0.021 & 0.120 & -0.021 & 0.118 & -0.021 & 0.118 \\\cmidrule{2-11}
  & 0.5 & $\beta_{p0}$ & -0.056 & 0.193 & -0.056 & 0.192 & -0.056 & 0.183 & -0.056 & 0.184 \\
  &     & $\beta_{p1}$ &  0.013 & 0.117 &  0.012 & 0.117 &  0.012 & 0.120 &  0.011 & 0.119 \\
  &     & $\delta_p$   & -0.006 & 0.072 & -0.006 & 0.072 & -0.006 & 0.072 & -0.006 & 0.072 \\\hline
4 & 0.1 & $\beta_{p0}$ &  0.053 & 0.246 &  0.054 & 0.246 &  0.063 & 0.237 &  0.063 & 0.236 \\
  &     & $\beta_{p1}$ &  0.002 & 0.152 &  0.002 & 0.150 &  0.004 & 0.143 &  0.003 & 0.145 \\
  &     & $\delta_p$   & -0.018 & 0.102 & -0.017 & 0.101 & -0.017 & 0.097 & -0.016 & 0.098 \\\cmidrule{2-11}
  & 0.5 & $\beta_{p0}$ &  0.003 & 0.202 &  0.003 & 0.203 &  0.015 & 0.193 &  0.013 & 0.191 \\
  &     & $\beta_{p1}$ & -0.003 & 0.140 & -0.002 & 0.143 & -0.003 & 0.130 & -0.002 & 0.129 \\
  &     & $\delta_p$   & -0.009 & 0.096 & -0.010 & 0.098 & -0.007 & 0.086 & -0.007 & 0.086 \\\hline
5 & 0.1 & $\beta_{p0}$ &  0.011 & 0.300 &  0.010 & 0.301 & -0.010 & 0.296 & -0.014 & 0.295 \\
  &     & $\beta_{p1}$ &  0.030 & 0.200 &  0.031 & 0.199 &  0.032 & 0.207 &  0.031 & 0.206 \\
  &     & $\delta_p$   & -0.050 & 0.148 & -0.050 & 0.147 & -0.054 & 0.150 & -0.054 & 0.150 \\\cmidrule{2-11}
  & 0.5 & $\beta_{p0}$ & -0.046 & 0.206 & -0.046 & 0.206 & -0.065 & 0.202 & -0.073 & 0.206 \\
  &     & $\beta_{p1}$ &  0.052 & 0.148 &  0.053 & 0.148 &  0.053 & 0.154 &  0.055 & 0.156 \\
  &     & $\delta_p$   & -0.042 & 0.101 & -0.043 & 0.102 & -0.046 & 0.106 & -0.046 & 0.107 \\\bottomrule
\end{tabular}

}

\end{table}

Next, the two alternative prior specifications for $\eta_p$, $\sigma$, and $\phi$ are considered to study the prior sensitivity. 
The first alternative specification considers the more diffuse priors given by $\eta_p\sim\N(0,25)$, $\sigma\sim\IG(0.1,0.1)$, and $\phi\sim\IG(0.01,0.01)$. 
The second alternative specification is the even more diffuse setting given by $\eta_p\sim\N(0,100)$, $\sigma\sim\IG(0.001,0.001)$, and $\phi\sim\IG(0.001,0.001)$. 
For ALDP and SNDP, the default base measures are used. 
For $\vbeta_p$, $\delta_p$, and $\vgamma$, we use the default specification. 
Table~\ref{tab:prior} presents the biases and RMSEs for AL, SN, ALDP, and SNDP under the five simulation settings for $p=0.1$ and $0.5$, showing that the result is robust with respect to the choice of hyperparameters. 
We also considered some different prior choices for $(\vbeta_p',\delta_p)'$ and $\vgamma$, and obtained robust results.

\renewcommand{\baselinestretch}{1.0}
\renewcommand{\arraystretch}{1.0}
\renewcommand{\tabcolsep}{0.5mm}
\begin{table}[H]
\caption{Biases and RMSEs under the alternative priors for $\sigma$, $\tau$, and $\eta_p$}
\label{tab:prior}
\centering
{\scriptsize
\begin{tabular}{cccrrrrrrrrrrrrrrrr}\toprule
\multicolumn{3}{c}{}& \multicolumn{4}{c}{AL} & \multicolumn{4}{c}{SN} & \multicolumn{4}{c}{ALDP} & \multicolumn{4}{c}{SNDP} \\
\cmidrule(lr){4-7}\cmidrule(lr){8-11}\cmidrule(lr){12-15}\cmidrule(lr){16-19}
\multicolumn{3}{c}{}&\multicolumn{2}{c}{Alternative 1}&\multicolumn{2}{c}{Alternative 2}&\multicolumn{2}{c}{Alternative 1}&\multicolumn{2}{c}{Alternative 2}&\multicolumn{2}{c}{Alternative 1}&\multicolumn{2}{c}{Alternative 2}&\multicolumn{2}{c}{Alternative 1}&\multicolumn{2}{c}{Alternative 2}\\
\cmidrule(lr){4-5}\cmidrule(lr){6-7}\cmidrule(lr){8-9}\cmidrule(lr){10-11}\cmidrule(lr){12-13}\cmidrule(lr){14-15}
\cmidrule(lr){16-17}\cmidrule(lr){18-19}
Setting & $p$ & Parameter  & \multicolumn{1}{c}{Bias} & \multicolumn{1}{c}{RMSE}  & \multicolumn{1}{c}{Bias} & \multicolumn{1}{c}{RMSE}  & \multicolumn{1}{c}{Bias} & \multicolumn{1}{c}{RMSE}  & \multicolumn{1}{c}{Bias} & \multicolumn{1}{c}{RMSE}  & \multicolumn{1}{c}{Bias} & \multicolumn{1}{c}{RMSE}  & \multicolumn{1}{c}{Bias} & \multicolumn{1}{c}{RMSE}  & \multicolumn{1}{c}{Bias} & \multicolumn{1}{c}{RMSE}  & \multicolumn{1}{c}{Bias} & \multicolumn{1}{c}{RMSE} \\\hline
1 & 0.1 & $\beta_{p0}$ &  0.049 & 0.239 &  0.050 & 0.240 &  0.050 & 0.212 &  0.050 & 0.211 &  0.050 & 0.237 &  0.049 & 0.239 &  0.051 & 0.211 &  0.051 & 0.211 \\
  &     & $\beta_{p1}$ & -0.022 & 0.140 & -0.021 & 0.140 & -0.019 & 0.135 & -0.019 & 0.135 & -0.022 & 0.139 & -0.022 & 0.139 & -0.021 & 0.135 & -0.020 & 0.135 \\
  &     & $\delta_p$   & -0.009 & 0.092 & -0.009 & 0.092 & -0.008 & 0.085 & -0.008 & 0.085 & -0.009 & 0.092 & -0.009 & 0.093 & -0.008 & 0.085 & -0.008 & 0.085 \\\cmidrule{2-19}
  & 0.5 & $\beta_{p0}$ &  0.018 & 0.180 &  0.017 & 0.181 &  0.018 & 0.166 &  0.019 & 0.168 &  0.018 & 0.183 &  0.016 & 0.180 &  0.019 & 0.165 &  0.018 & 0.165 \\
  &     & $\beta_{p1}$ & -0.001 & 0.089 & -0.001 & 0.089 &  0.001 & 0.087 &  0.002 & 0.087 & -0.001 & 0.089 & -0.001 & 0.089 &  0.001 & 0.087 &  0.001 & 0.087 \\
  &     & $\delta_p$   & -0.004 & 0.063 & -0.004 & 0.063 & -0.003 & 0.061 & -0.003 & 0.061 & -0.004 & 0.063 & -0.004 & 0.063 & -0.003 & 0.061 & -0.003 & 0.061 \\\hline
2 & 0.1 & $\beta_{p0}$ &  0.091 & 0.304 &  0.088 & 0.304 &  0.084 & 0.312 &  0.081 & 0.311 &  0.092 & 0.305 &  0.090 & 0.307 &  0.094 & 0.300 &  0.097 & 0.305 \\
  &     & $\beta_{p1}$ & -0.008 & 0.160 & -0.008 & 0.160 & -0.008 & 0.164 & -0.008 & 0.164 & -0.007 & 0.159 & -0.007 & 0.160 & -0.010 & 0.159 & -0.010 & 0.159 \\
  &     & $\delta_p$   & -0.025 & 0.116 & -0.025 & 0.115 & -0.024 & 0.116 & -0.023 & 0.115 & -0.025 & 0.115 & -0.025 & 0.116 & -0.023 & 0.113 & -0.024 & 0.114 \\\cmidrule{2-19}
  & 0.5 & $\beta_{p0}$ & -0.001 & 0.198 & -0.001 & 0.199 & -0.011 & 0.207 & -0.011 & 0.209 &  0.002 & 0.199 &  0.001 & 0.200 &  0.005 & 0.196 &  0.005 & 0.195 \\
  &     & $\beta_{p1}$ &  0.003 & 0.126 &  0.003 & 0.126 &  0.003 & 0.130 &  0.002 & 0.129 &  0.003 & 0.126 &  0.003 & 0.127 & -0.000 & 0.123 &  0.000 & 0.123 \\
  &     & $\delta_p$   & -0.004 & 0.082 & -0.004 & 0.082 & -0.002 & 0.084 & -0.002 & 0.083 & -0.004 & 0.082 & -0.004 & 0.082 & -0.002 & 0.081 & -0.002 & 0.081 \\\hline
3 & 0.1 & $\beta_{p0}$ &  0.027 & 0.284 &  0.028 & 0.287 & -0.004 & 0.301 & -0.006 & 0.301 &  0.028 & 0.286 &  0.027 & 0.287 &  0.024 & 0.282 &  0.026 & 0.282 \\
  &     & $\beta_{p1}$ & -0.008 & 0.180 & -0.007 & 0.180 & -0.007 & 0.182 & -0.007 & 0.182 & -0.008 & 0.180 & -0.008 & 0.180 & -0.009 & 0.181 & -0.009 & 0.182 \\
  &     & $\delta_p$   & -0.022 & 0.120 & -0.022 & 0.120 & -0.022 & 0.118 & -0.022 & 0.118 & -0.021 & 0.120 & -0.022 & 0.120 & -0.021 & 0.119 & -0.021 & 0.118 \\\cmidrule{2-19}
  & 0.5 & $\beta_{p0}$ & -0.054 & 0.191 & -0.053 & 0.191 & -0.083 & 0.223 & -0.084 & 0.223 & -0.054 & 0.193 & -0.056 & 0.192 & -0.055 & 0.182 & -0.054 & 0.183 \\
  &     & $\beta_{p1}$ &  0.013 & 0.117 &  0.014 & 0.117 &  0.014 & 0.122 &  0.015 & 0.123 &  0.013 & 0.117 &  0.013 & 0.117 &  0.012 & 0.119 &  0.012 & 0.120 \\
  &     & $\delta_p$   & -0.006 & 0.072 & -0.006 & 0.072 & -0.008 & 0.071 & -0.007 & 0.071 & -0.007 & 0.072 & -0.006 & 0.072 & -0.006 & 0.072 & -0.007 & 0.072 \\\hline
4 & 0.1 & $\beta_{p0}$ &  0.056 & 0.245 &  0.055 & 0.246 &  0.066 & 0.232 &  0.066 & 0.231 &  0.054 & 0.245 &  0.054 & 0.247 &  0.063 & 0.235 &  0.063 & 0.233 \\
  &     & $\beta_{p1}$ &  0.004 & 0.152 &  0.002 & 0.151 &  0.005 & 0.142 &  0.006 & 0.143 &  0.003 & 0.151 &  0.003 & 0.150 &  0.004 & 0.143 &  0.005 & 0.145 \\
  &     & $\delta_p$   & -0.019 & 0.103 & -0.018 & 0.102 & -0.019 & 0.100 & -0.020 & 0.101 & -0.018 & 0.102 & -0.018 & 0.101 & -0.017 & 0.098 & -0.018 & 0.099 \\\cmidrule{2-19}
  & 0.5 & $\beta_{p0}$ &  0.006 & 0.198 &  0.007 & 0.200 &  0.017 & 0.186 &  0.020 & 0.189 &  0.006 & 0.202 &  0.005 & 0.202 &  0.017 & 0.191 &  0.016 & 0.191 \\
  &     & $\beta_{p1}$ & -0.003 & 0.141 & -0.003 & 0.142 &  0.001 & 0.129 &  0.001 & 0.129 & -0.002 & 0.143 & -0.002 & 0.143 & -0.001 & 0.131 & -0.002 & 0.129 \\
  &     & $\delta_p$   & -0.011 & 0.097 & -0.010 & 0.097 & -0.011 & 0.089 & -0.011 & 0.089 & -0.011 & 0.099 & -0.010 & 0.097 & -0.009 & 0.087 & -0.008 & 0.085 \\\hline
5 & 0.1 & $\beta_{p0}$ &  0.017 & 0.299 &  0.017 & 0.301 & -0.029 & 0.342 & -0.028 & 0.340 &  0.018 & 0.302 &  0.014 & 0.300 & -0.003 & 0.296 & -0.003 & 0.294 \\
  &     & $\beta_{p1}$ &  0.030 & 0.199 &  0.030 & 0.201 &  0.045 & 0.223 &  0.045 & 0.224 &  0.031 & 0.200 &  0.030 & 0.198 &  0.032 & 0.207 &  0.032 & 0.207 \\
  &     & $\delta_p$   & -0.050 & 0.147 & -0.050 & 0.148 & -0.067 & 0.161 & -0.068 & 0.164 & -0.051 & 0.149 & -0.050 & 0.148 & -0.053 & 0.149 & -0.052 & 0.148 \\\cmidrule{2-19}
  & 0.5 & $\beta_{p0}$ & -0.041 & 0.199 & -0.041 & 0.201 & -0.085 & 0.256 & -0.087 & 0.256 & -0.043 & 0.206 & -0.042 & 0.206 & -0.059 & 0.199 & -0.061 & 0.201 \\
  &     & $\beta_{p1}$ &  0.053 & 0.148 &  0.053 & 0.149 &  0.069 & 0.181 &  0.070 & 0.182 &  0.053 & 0.148 &  0.053 & 0.148 &  0.054 & 0.155 &  0.054 & 0.155 \\
  &     & $\delta_p$   & -0.042 & 0.102 & -0.042 & 0.102 & -0.059 & 0.123 & -0.059 & 0.124 & -0.042 & 0.102 & -0.043 & 0.103 & -0.045 & 0.106 & -0.046 & 0.107 \\\bottomrule
\end{tabular}
}
\end{table}

These findings thus confirm the robustness of the results with respect to the choice of base measures and prior distributions provided that a valid instrument is available. 
In the context of mean regression models, however, when the instrument is weak, the posterior distribution is known to exhibit sharp behaviour in the vicinity of non-identifiability (Hoogerheide~\etal,~2007b) and the posterior distribution is greatly affected by the prior specification (\eg Lopes and Polson,~2014). 

Here, we illustrate the behaviour of the posterior distribution by using a weak instrument. 
The data are generated from (\ref{eqn:sim1}) without the regressor: 
\begin{equation}\label{eqn:sim2}
\begin{split}
y_i^*&=\delta d_i + \eta v_i + e_i,\\
d_i &= \gamma w_i + v_i, 
\end{split}
\end{equation}
for $i=1,\dots,300$, where $\gamma=0.1$, $(\delta,\eta)=(1,0.6)$, $w_i\sim\N(0,1)$, $v_i\sim\N(0,1)$, and $e_i\sim\N(0,1-\eta^2)$. 
The AL and SN models are estimated for $p=0.1$ by running the MCMC for 20000 iterations and discarding the first 5000 draws as the burn-in period under the three prior specifications previously considered. 

\begin{figure}[H]
\centering
\includegraphics[width=\textwidth]{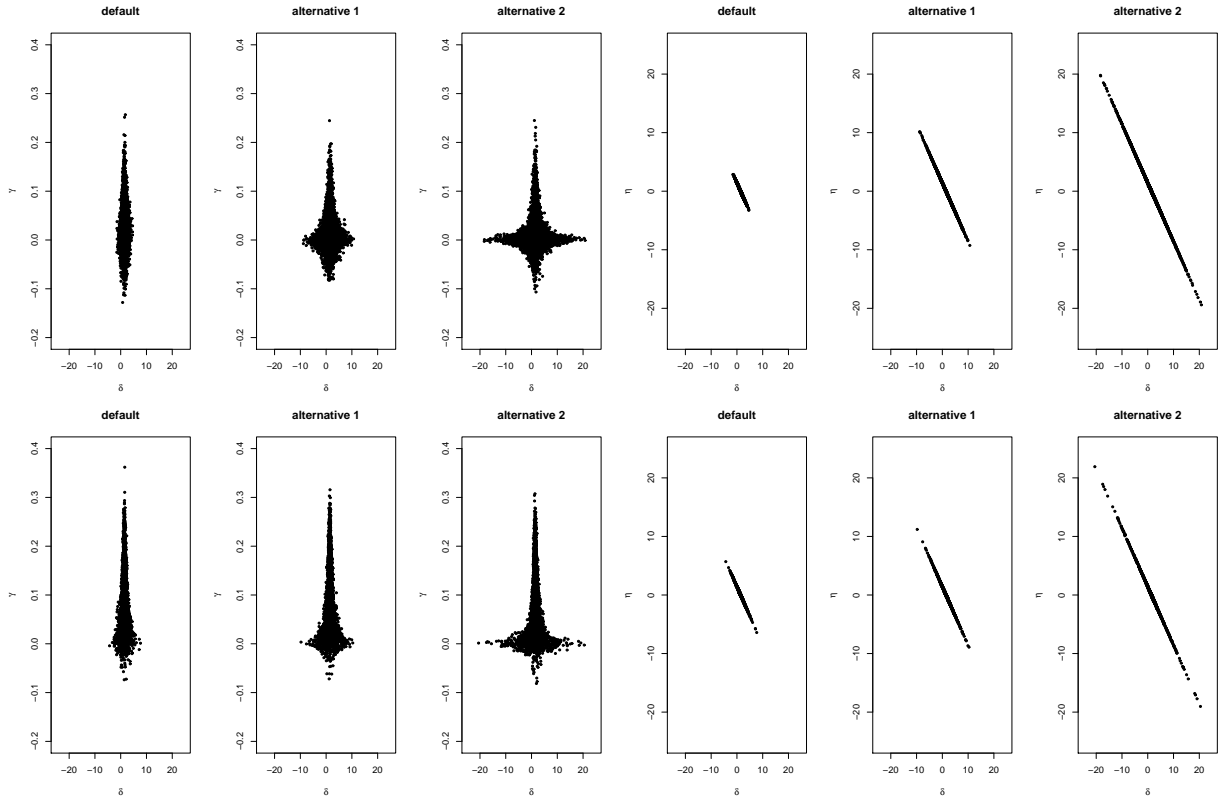}
\caption{Joint posterior of $(\delta,\gamma)$ and $(\delta,\eta)$ for AL (top row) and SN (bottom row)}
\label{fig:fig2}
\end{figure}

Figure~\ref{fig:fig2} presents the joint posterior distribution of $(\delta,\gamma)$ and $(\delta,\eta)$ for AL and SN under the three prior specifications and shows that the posterior distribution is greatly affected by the prior specification. 
The posterior distribution of $\delta$ becomes more diffuse as $\gamma$ approaches zero. 
This trend becomes more profound as we use more diffuse prior distributions, producing star shapes. 
The figure also suggests that the prior distribution can act as an informative prior about the linear relationship between $\delta$ and $\eta$. 
Similar results were also obtained under different prior specifications for $\vbeta_p$, $\delta_p$, and $\vgamma$ as well as for ALDP and SNDP.

\section{Application: Labour Force Participation of Married Women}\label{sec:real}
The proposed endogenous models are applied to the dataset on the labour supply of married women of Mroz~(1987). 
The dataset includes observations on $753$ individuals. 
The response variable is the total number of hours in every $100$ hours the wife worked for a wage outside the home during 1975. 
In the data, 325 of the 753 women worked zero hours and the corresponding responses are treated as left censored at zero. 
Hence, the censoring rate is approximately $0.43$. 
The regressors of our model include years of education (\textit{educ}), years of experience (\textit{exper}) and its square (\textit{expersq}), age of the wife (\textit{age}), number of children under 6 years old (\textit{kidslt6}), number of children equal to or greater than 6 years old (\textit{kidsge6}), and non-wife household income (\textit{nwifeinc}). 
We treat \textit{nwifeinc} as an endogenous variable because it may be correlated with the unobserved household preference for the labour force participation of the wife. 
As an instrument, we include the years of education of the husband (\textit{huseduc}), since this can influence both his income and the non-wife household income, but it should not influence the decision of the wife to participate in the labour force. 
Smith and Blundell~(1986) considered a similar setting where non-wife income was considered to be endogenous and the education of the husband was employed as the instrumental variable. 
They applied the endogenous Tobit model to data derived from the 1981 Family Expenditure Survey in the United Kingdom. 

\begin{figure}[H]
\centering
\includegraphics[width=\textwidth]{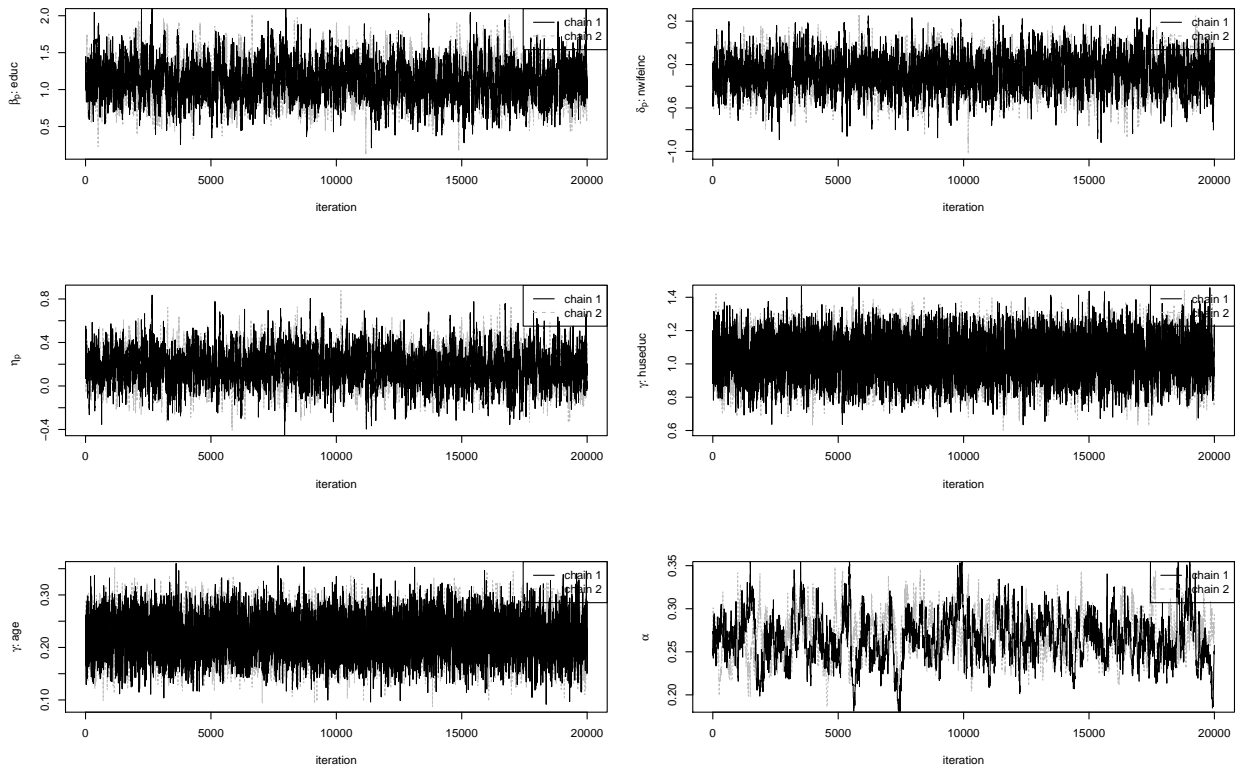}
\caption{Post burn-in trace plots for SNDP for $p=0.1$}
\label{fig:fig3}
\end{figure}

Using the default prior specifications, the ALDP and SNDP models are estimated for $p=0.05,0.1,\dots,0.95$ by running the MCMC for $30000$ iterations and discarding the first $10000$ draws as the burn-in period.  
Convergence is monitored by using the trace plots and Gelman-Rubin statistic for two chains with widespread starting values (Gelman~\etal,~2014). 
The upper bounds of the Gelman-Rubin  confidence intervals for the selected parameters, $\beta_{p,\textit{educ}}$, $\delta_p$, $\eta_p$, $\gamma_{\textit{huseduc}}$, $\gamma_{\textit{age}}$, and $\alpha$, for SNDP in the case of $p=0.1$ are $1.01$, $1.01$, $1.01$, $1.00$, $1.00$, and $1.06$ , respectively. 
Figure~\ref{fig:fig3} presents the post burn-in trace plots for these parameters and shows the evidence of convergence of the chains.

First, we present the results for the representative quantiles, $p=0.1$, $0.5$, and $0.9$.  
Table~\ref{tab:real} shows the posterior means, 95\% credible intervals, and inefficiency factors for ALDP and SNDP for these quantiles. 
The table shows that the sampling algorithm worked efficiently as the inefficiency factors are reasonably small. 
The posterior means for the instrument, \textit{huseduc}, are positive and the 95\% credible intervals do not include zero for all cases for both models, implying that \textit{huseduc} is a valid instrument. 
For $p=0.5$, the posterior means for $\eta_p$ are $0.450$ and $0.446$ for ALDP and SNDP, respectively, and the 95\% credible intervals do not include zero. 
Therefore, it is suggested that non-wife income be treated as an endogenous variable for the median regression.

\renewcommand{\baselinestretch}{1.0}
\renewcommand{\arraystretch}{1.0}
\renewcommand{\tabcolsep}{0.5mm}
\begin{table}[H]
\caption{Posterior Summary for Female Labour Data}
\label{tab:real}
\centering
{\scriptsize
\begin{tabular}{cclr@{\quad[}rr@{]\quad}rr@{\quad[}rr@{]\quad}r}\toprule
&&&\multicolumn{4}{c}{ALDP}&\multicolumn{4}{c}{SNDP}\\
\cmidrule(lr){4-7}\cmidrule(lr){8-11}
$p$ & \multicolumn{2}{l}{Parameter} & \multicolumn{1}{c}{Mean} & \multicolumn{2}{c}{95\% CI} & \multicolumn{1}{c}{IF} & \multicolumn{1}{c}{Mean} & \multicolumn{2}{c}{95\% CI} & \multicolumn{1}{c}{IF}  \\\hline
0.1 & $\vbeta_p$ & constant          &  -4.205  & -10.758, &   2.430 & 13.0 &  -4.340  & -11.121, &  2.288 & 18.2 \\
    &            & \textit{educ}     &   1.126  &   0.656, &   1.599 & 12.7 &   1.117  &   0.659, &  1.614 & 47.7 \\
    &            & \textit{age}      &  -0.436  &  -0.565, &  -0.311 & 25.1 &  -0.424  &  -0.554, & -0.293 & 37.0 \\
    &            & \textit{exper}    &   1.070  &   0.731, &   1.437 & 41.3 &   1.051  &   0.723, &  1.387 & 92.3 \\
    &            & \textit{expersq}  &  -0.019  &  -0.030, &  -0.009 & 24.7 &  -0.019  &  -0.029, & -0.009 & 52.4 \\
    &            & \textit{kidslt6}  &  -8.346  & -11.145, &  -5.949 & 80.4 &  -8.296  & -10.948, & -5.861 & 65.6 \\
    &            & \textit{kigsge6}  &   0.068  &  -0.487, &   0.534 & 31.7 &   0.045  &  -0.528, &  0.512 & 14.4 \\
    & $\delta_p$ & \textit{nwifeinc} &  -0.284  &  -0.584, &   0.010 & 14.1 &  -0.279  &  -0.577, &  0.007 & 33.4 \\
    & $\eta_p$   &                   &   0.176  &  -0.117, &   0.473 & 11.5 &   0.171  &  -0.125, &  0.472 & 27.8 \\\cmidrule{2-11}
    & $\vgamma$  & constant          & -10.117  & -14.486, &  -5.490 &  9.4 & -10.609  & -15.023, & -6.112 & 11.3 \\
    &            & \textit{huseduc}  &   1.013  &   0.771, &   1.239 &  9.7 &   1.037  &   0.812, &  1.257 &  4.9 \\
    &            & \textit{educ}     &   0.272  &   0.018, &   0.551 &  6.6 &   0.286  &   0.018, &  0.562 &  5.7 \\
    &            & \textit{age}      &   0.210  &   0.140, &   0.280 &  6.7 &   0.221  &   0.152, &  0.290 &  7.6 \\
    &            & \textit{exper}    &  -0.090  &  -0.269, &   0.084 & 12.2 &  -0.122  &  -0.301, &  0.052 &  8.5 \\
    &            & \textit{expersq}  &  -0.003  &  -0.009, &   0.003 & 12.2 &  -0.002  &  -0.008, &  0.003 &  4.7 \\
    &            & \textit{kidslt6}  &  -0.554  &  -1.424, &   0.351 &  6.4 &  -0.472  &  -1.430, &  0.469 &  5.5 \\
    &            & \textit{kigsge6}  &   0.481  &   0.125, &   0.838 &  6.9 &   0.464  &   0.080, &  0.839 &  7.7 \\
    & $\alpha$   &                   &   0.250  &   0.211, &   0.298 & 33.9 &   0.265  &   0.212, &  0.322 & 78.7 \\\hline
0.5 & $\vbeta_p$ & constant          &   8.571  &  -0.899, &  17.634 &  7.1 &   8.265  &  -1.288, & 17.473 &  9.1 \\
    &            & \textit{educ}     &   1.287  &   0.734, &   1.889 & 11.0 &   1.291  &   0.727, &  1.895 &  8.2 \\
    &            & \textit{age}      &  -0.510  &  -0.680, &  -0.333 & 10.9 &  -0.502  &  -0.670, & -0.321 & 12.9 \\
    &            & \textit{exper}    &   1.398  &   1.029, &   1.787 & 12.0 &   1.391  &   1.021, &  1.777 & 12.1 \\
    &            & \textit{expersq}  &  -0.021  &  -0.034, &  -0.009 & 13.4 &  -0.021  &  -0.034, & -0.009 & 13.3 \\
    &            & \textit{kidslt6}  &  -9.546  & -11.975, &  -7.305 & 14.5 &  -9.441  & -11.849, & -7.123 &  5.2 \\
    &            & \textit{kigsge6}  &  -0.255  &  -1.116, &   0.620 & 10.9 &  -0.268  &  -1.104, &  0.592 & 10.4 \\
    & $\delta_p$ & \textit{nwifeinc} &  -0.525  &  -0.944, &  -0.159 & 15.5 &  -0.522  &  -0.917, & -0.165 &  8.5 \\
    & $\eta_p$   &                   &   0.450  &   0.079, &   0.885 & 14.0 &   0.446  &   0.087, &  0.852 &  7.9 \\\cmidrule{2-11}
    & $\vgamma$  & constant          & -10.318  & -14.784, &  -5.689 & 12.3 & -11.021  & -15.556, & -6.377 & 11.2 \\
    &            & \textit{huseduc}  &   1.013  &   0.768, &   1.242 &  9.7 &   1.032  &   0.809, &  1.251 &  7.7 \\
    &            & \textit{educ}     &   0.277  &   0.025, &   0.557 &  5.8 &   0.301  &   0.028, &  0.583 &  5.5 \\
    &            & \textit{age}      &   0.212  &   0.142, &   0.283 &  8.2 &   0.226  &   0.156, &  0.296 & 11.5 \\
    &            & \textit{exper}    &  -0.090  &  -0.274, &   0.084 &  4.3 &  -0.120  &  -0.298, &  0.054 &  9.4 \\
    &            & \textit{expersq}  &  -0.003  &  -0.009, &   0.003 &  4.4 &  -0.002  &  -0.008, &  0.003 &  8.4 \\
    &            & \textit{kidslt6}  &  -0.536  &  -1.408, &   0.362 &  2.9 &  -0.447  &  -1.413, &  0.515 &  3.1 \\
    &            & \textit{kigsge6}  &   0.491  &   0.136, &   0.850 &  2.7 &   0.468  &   0.082, &  0.863 &  4.9 \\
    & $\alpha$   &                   &   0.250  &   0.212, &   0.297 & 18.6 &   0.263  &   0.215, &  0.315 & 77.0 \\\hline
0.9 & $\vbeta_p$ & constant          &  17.077  &   9.225, &  25.430 &  7.5 &  16.957  &   8.985, & 25.429 &  3.2 \\
    &            & \textit{educ}     &   0.405  &  -0.107, &   0.905 &  7.9 &   0.420  &  -0.102, &  0.921 &  2.1 \\
    &            & \textit{age}      &  -0.266  &  -0.424, &  -0.112 &  6.7 &  -0.265  &  -0.419, & -0.113 &  2.7 \\
    &            & \textit{exper}    &   1.075  &   0.749, &   1.387 & 13.1 &   1.072  &   0.747, &  1.389 & 12.0 \\
    &            & \textit{expersq}  &  -0.018  &  -0.026, &  -0.010 & 10.8 &  -0.018  &  -0.026, & -0.010 &  8.8 \\
    &            & \textit{kidslt6}  &  -6.014  &  -8.373, &  -3.553 &  7.9 &  -6.085  &  -8.476, & -3.584 &  8.3 \\
    &            & \textit{kigsge6}  &   0.254  &  -0.490, &   0.978 &  5.7 &   0.254  &  -0.492, &  1.009 &  9.9 \\
    & $\delta_p$ & \textit{nwifeinc} &  -0.043  &  -0.384, &   0.288 &  4.5 &  -0.050  &  -0.380, &  0.275 &  6.7 \\
    & $\eta_p$   &                   &  -0.002  &  -0.340, &   0.339 &  5.0 &   0.004  &  -0.328, &  0.337 &  6.8 \\\cmidrule{2-11}
    & $\vgamma$  & constant          & -10.174  & -14.698, &  -5.486 &  6.1 & -10.741  & -15.390, & -5.946 & 16.1 \\
    &            & \textit{huseduc}  &   1.013  &   0.773, &   1.240 &  9.2 &   1.036  &   0.812, &  1.253 &  6.3 \\
    &            & \textit{educ}     &   0.274  &   0.021, &   0.551 &  9.5 &   0.292  &   0.017, &  0.582 &  9.5 \\
    &            & \textit{age}      &   0.211  &   0.138, &   0.283 &  4.2 &   0.223  &   0.151, &  0.294 & 14.0 \\
    &            & \textit{exper}    &  -0.092  &  -0.272, &   0.085 &  5.8 &  -0.126  &  -0.300, &  0.048 &  6.9 \\
    &            & \textit{expersq}  &  -0.003  &  -0.009, &   0.003 &  7.0 &  -0.002  &  -0.008, &  0.003 &  6.0 \\
    &            & \textit{kidslt6}  &  -0.550  &  -1.435, &   0.349 &  7.2 &  -0.483  &  -1.450, &  0.500 &  9.5 \\
    &            & \textit{kigsge6}  &   0.483  &   0.128, &   0.837 &  4.3 &   0.464  &   0.076, &  0.857 &  4.2 \\
    & $\alpha$   &                   &   0.251  &   0.213, &   0.291 & 34.5 &   0.265  &   0.213, &  0.321 & 66.1 \\\bottomrule
\end{tabular}
}
\end{table}

\begin{figure}[H]
\centering
\includegraphics[width=\textwidth]{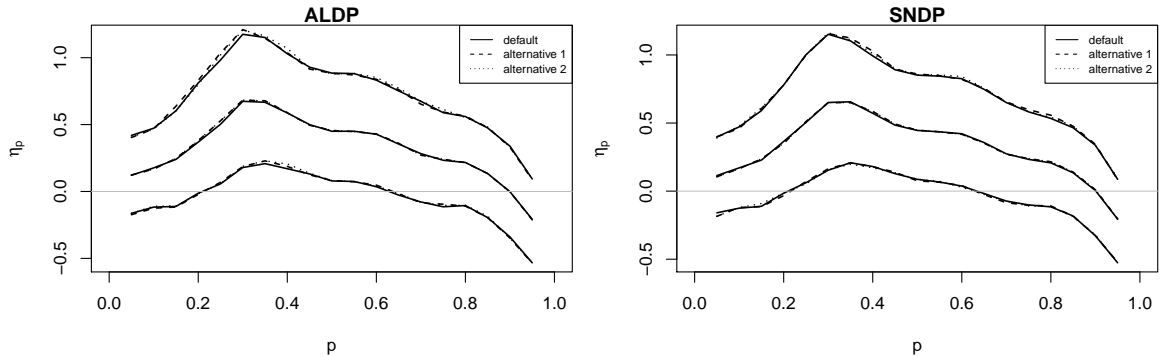}
\caption{Posterior means and 95\% credible intervals of $\eta_p$ under the default and alternative priors for $p=0.05,0.1,\dots,0.95$}
\label{fig:fig4}
\end{figure}

To study the endogeneity in non-wife household income across quantiles, the posterior distributions of $\eta_p$ are presented. 
The results across the quantiles can be best understood by plotting the posterior distributions as a function of $p$. 
Figure~\ref{fig:fig4} shows the posterior means and 95\% credible intervals of $\eta_p$ for ALDP and SNDP for $p=0.05,0.1,\dots,0.95$. 
The figure shows that the two models produced similar results and that the posterior distributions of $\eta_p$ are concentrated away from zero for the mid quantiles. 
Specifically, for $0.2<p<0.65$, the 95\% credible intervals do not include zero for either model. 
There are notable peaks around $p=0.35$, where the posterior means of $\eta_p$ under the default prior specifications are $0.664$ and $0.662$ with the 95\% credible intervals $(0.201,1.137)$ and $(0.230,1.124)$ for ALDP and SNDP, respectively. 
This is an interesting result considering that the censoring rate is $0.43$. 
The result implies that the effect of the endogeneity of non-wife income is the most profound when the wife is about to decide whether to enter the labour force.  
When the opportunity cost of labour supply is very high (lower quantile) or the wife works on a more regular basis (higher quantile), such endogeneity diminishes. 
Smith and Blundell~(1986) also reported that non-wife income is endogenous by using the endogenous Tobit regression model. 
The mean of our dataset is $7.399$, which approximately corresponds to the $0.6$-th quantile. 
For $p=0.6$, the posterior mean of $\eta_p$ for ALDP is $0.428$ with the 95\% credible interval $(0.036,0.832)$ and that for SNDP is $0.421$ with the 95\% credible interval $(0.037, 0.832)$. 
The figure also shows the posterior means and 95\% credible intervals under the two alternative prior specifications considered in Section~\ref{sec:alt}, confirming that our results are robust with respect to the prior specifications.

\begin{figure}[H]
\centering
\includegraphics[width=\textwidth]{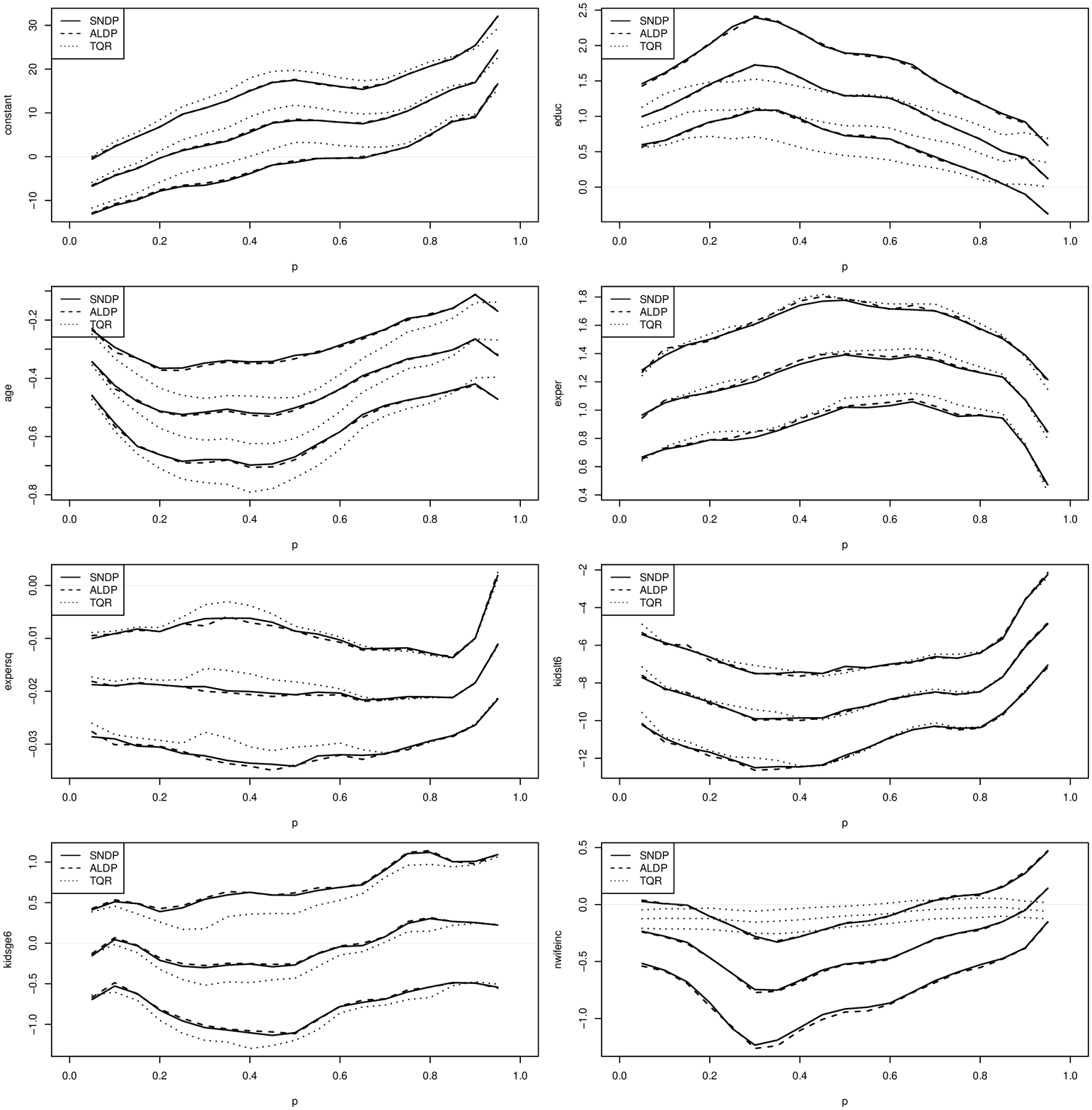}
\caption{Posterior means and 95\% credible intervals of $(\vbeta_p',\delta_p)'$ for ALDP, SNDP, and TQR for  $p=0.05,0.1,\dots,0.95$}
\label{fig:fig5}
\end{figure}

Figure~\ref{fig:fig5} compares the posterior means and 95\% credible intervals of $(\vbeta_p',\delta_p)'$ for SNDP, ALDP, and TQR for $p=0.05,0.1,\dots,0.95$. 
The results for SNDP and ALDP are quite similar. 
The figure clearly shows that the posterior distributions for the key variable, \textit{nwifeinc}, for the proposed models and TQR exhibit some differences for $0.2<p<0.65$, where \textit{nwifeinc} is indicated to be endogenous. 
The difference becomes the most profound around $p=0.35$ for which the posterior mean for \textit{nwifeinc} is $-0.761$ for ALDP, $-0.756$ for SNDP, and $-0.147$ for TQR, implying a stronger effect of non-wife income when endogeneity is taken into account. 
The posterior distributions for \textit{nwifeinc} for ALDP and SNDP are more dispersed than that for TQR for all $p$. 
While the 95\% credible intervals include zero for all models for the upper quantiles, for the lower quantiles, such as $p=0.1$, those for ALDP and SNDP include zero and those for TQR do not.

Differences in the results are also observed for other variables. 
For $p=0.35$, the posterior means for \textit{educ} and \textit{age} are respectively $1.689$ and $-0.513$ for ALDP,  $1.705$ and $-0.504$ for SNDP,  and $1.064$ and $-0.606$ for TQR. 
For the upper quantiles, $p>0.85$, the 95\% credible intervals for \textit{educ} include zero for the proposed models, while those for TQR do not, implying that an additional year of education does not increase the working hours for those quantiles when the endogeneity from non-wife income is taken into account. 
For \textit{expersq}, the endogenous models result in slightly more dispersed posterior distributions for $0.2<p<0.7$. 
The posterior means for $p=0.35$ are $-0.021$, $-0.020$, and $-0.016$ for ALDP, SNDP, and TQR, respectively. 
For \textit{kidsge6}, the posterior means for $p=0.35$ are $-0.274$, $-0.262$, and $-0.475$ for ALDP, SNDP, and TQR, respectively. 
However, the 95\% credible intervals include zero for all $p$ for all models. 
On the other hand, the figure also shows that the models produced similar results for \textit{exper} and \textit{kidslt6} for all $p$.

\section{Conclusion}\label{sec:conc}
We proposed Bayesian endogenous TQR models using parametric and semiparametric first stage regression models built around the zero $\alpha$-th quantile assumption. 
The value of $\alpha$ determines the quantile level of the mode of the error distribution and is estimated from the data. 
From the simulation study, the AL, ALDP, and SNDP models worked relatively well for the various situations, while they faced the same limitation pointed out by Kottas and Krnjaji\'c~(2011). 
On the other hand, the SN model could not accommodate the fat tailed first stage errors. 
Although AEP could be a promising model in terms of flexibility, the inefficiency of the MCMC algorithm largely limits its applicability in practice. 
The development of a more convenient mixture representation for the AEP distribution is thus required. 
From application to data on the labour supply of married women, the effect of the endogeneity in non-wife income was found to be the most profound for the quantile level close to the censoring rate.
For this quantile, some differences in the parameter estimates between the endogenous and standard models were found, such as the stronger effect of non-wife income on working hours.

This study only considered the case of continuous endogenous variables. 
We are also interested in incorporating endogenous binary variables into a Bayesian quantile regression model. 
An important extension might therefore be addressing multiple endogenous dummy variables to represent selection among multiple alternatives, such as the choice of a hospital and insurance plan, as considered in Geweke~\etal~(2003) and Deb~\etal~(2006). 
However, such an extension would be challenging with respect to the assumptions that must be imposed on the multivariate error terms. 
We leave these issues to future research.

\subsection*{Acknowledgements}
The author would like to thank the seminar participants at the second BAYSM and ESOBE~2014 and the anonymous referees for the valuable comments to improve the manuscript.  
The computational results were obtained by using Ox version 6.21 (Doornik,~2007). 
This study was supported by JSPS KAKENHI Grant Numbers 25245035, 26245028, 26380266, and 15K17036.


\end{document}